\documentclass[10pt,emulateapj,apj]{emulateapj}
\newcommand{\be}{\begin{equation}}
\newcommand{\ee}{\end{equation}}
\newcommand{\bea}{\begin{eqnarray}}
\newcommand{\eea}{\end{eqnarray}}

\shorttitle{Internal Extinction}
\shortauthors{Cho \& Park}

\begin{document}
\title{Internal Extinction in the SDSS Late-Type Galaxies} 
\author{Jungyeon Cho}
\affil{Dept. of Astronomy and Space Science,
       Chungnam National University, Daejeon, Korea}
\and
\author{Changbom Park\altaffilmark{1}} 
\affil{Korea Institute for Advanced Study, Seoul, Korea}
\altaffiltext{1}{email: cbp@kias.re.kr}

\begin{abstract}
We study internal extinction of late-type galaxies in the Sloan Digital
Sky Survey.
We find that the degree of internal extinction depends on
both the concentration index $c$ and $K_s$-band absolute 
magnitude $M_K$.
We give simple fitting functions for internal extinction.
In particular, we present analytic formulae giving the extinction-corrected
magnitudes from the observed optical parameters. For example, 
the extinction-corrected $r$-band absolute magnitude can be obtained by 
$M_{r,0}=-20.77 +(-1+\sqrt{1+4\Delta (M_{r,obs}+20.77+4.93\Delta)})/2\Delta$,
where $\Delta = 0.236\{1.35(c-2.48)^2-1.14\}\log(a/b)$, $c=R_{90}/R_{50}$
is the the concentration index, and $a/b$ is the isophotal axis ratio of 
the 25 mag/arcsec$^2$ isophote in the $i$-band. The $1\sigma$ error in 
$M_{r,0}$ is $0.21{\rm log}(a/b)$.
The late-type galaxies with very different inclinations
are found to trace almost the same sequence in the 
$(u-r)$-$M_r$ diagram when our prescriptions for extinction correction
are applied.
We also find that $(u-r)$ color can be a third independent parameter
that determines the degree of internal extinction.
\end{abstract}

\section{Introduction}
Dust in spiral galaxies causes internal extinction. 
For example, due to dust, 
edge-on spiral galaxies in general look fainter than face-on spirals
with the same intrinsic luminosity in short wavelength optical bands,
in particular (see Fig. 12 of  Choi, Park, \& Vogeley 2007).
The study of internal extinction is important for determination of
distances and absolute magnitude of galaxies.
Giovanelli et al.~(1995) demonstrated that
inadequate treatment of internal extinction has strong effects on
the distances estimated using
the Tully-Fisher relation (Tully \& Fisher 1977).
Internal extinction also causes biased measurements of galaxy
star formation rates (see, for example, Bell \& Kennicutt 2001;
Sullivan et al. 2000) and affects determination of galaxy luminosity
function (Shao et al.~2007).
Earlier studies of internal extinction include
Giovanelli et al.~(1994, 1995), Tully et al. (1998), and
Masters, Giovanelli, \& Haynes (2003).
Recent works that contain relevant discussions on the topic are
Rocha et al.~(2008), Unterborn \& Ryden (2008), and Maller et al.~(2008).

The amount of obscuration by dust is larger when the
observing wavelength is shorter.
Therefore,
the effect of inclination is more pronounced in short-wavelength bands, such as
$u$ or $r$ band, and it may be much smaller in $K_s$ (2.17$\mu$m) band.
As a consequence, when the viewing angle changes for a given galaxy, 
$u$- or $r$-band magnitude changes while $K_s$-band magnitude
does not change much.
Therefore, $u-K_s$ or $r-K_s$ color tends to be larger 
when the galaxy is viewed more edge-on.

Internal extinction depends on many factors.
Perhaps, the most important factor is the amount of dust.
How dust is distributed can also be an important factor.
The study of extinction versus inclination will ultimately
reveal how dust is distributed and how much dust is contained
in spiral galaxies.
Then, what determines the amount and distribution of dust in spiral galaxies?

Earlier works have discussed dependence of internal extinction on
luminosity.
Giovanelli et al.~(1995) found that
the amount of internal extinction in $I$ band depends on 
the galaxy luminosity.
Tully et al.~(1998) also found a strong luminosity dependence
using a magnitude-limited samples drawn 
{}from the Ursa Major and Pisces Clusters.
Masters, Giovanelli, \& Haynes (2003) studied internal extinction in spiral galaxies in the
near infrared and also found a luminosity dependence.
On the other hand, there are suggestions that
internal extinction depends on
galaxy type (de Vaucouleurs et al. 1991; Han 1992).

In this paper, we study the internal extinction in SDSS late-type galaxies.
We investigate the dependence of the inclination effects on luminosity,
concentration index $c$, and $u-r$ color.
In \S2, we describe the data set used in this paper.
In \S3, we study dependence of internal extinction on
the concentration index $c$ and $K_s$-band luminosity separately.
In \S4, we measure dependence of internal extinction on
the concentration index $c$ and $K_s$-band luminosity simultaneously.
In \S5, we derive dependence of inclination effects on $r$-band luminosity.
The result in \S5 is useful when 
$K$-band magnitude is not available.
In \S6, we present dependence on $u-r$ color.
We give discussion in \S7 and conclusion in \S8.

\section{Data and Method}
In this study, we investigate how $u$- and $r$-band
 magnitudes behave as 
the inclination angle changes.
The physical parameters we consider are
$r$-band absolute magnitude $M_r$,
$u-r$ color, the axis ratio $a/b$, and the concentration index $c$.

The primary data set we use is
a subset of volume-limited SDSS (DR5) galaxy sample 
(the data set D1 in Choi et al. 2007).
The redshift of the galaxies is between 0.0250 and 0.04374
and the minimum $r$-band absolute magnitude, $M_r$, is $-18.0+5{\rm log}h$,
where $h$ is the Hubble constant divided by 100 km s$^{-1}$ Mpc$^{-1}$.
In this paper, we assume the Hubble constant is 75km/sec/Mpc.
Therefore, the absolute magnitude in this paper can be transformed to the
$h$-dependent form by
\begin{equation}
M_{\lambda}^{h} = M_{\lambda}^{h=0.75} + 5\log(h/0.75),
\end{equation}
where $M_{\lambda}^{h=0.75}$ is the absolute magnitude we use in this paper.
The rest-frame absolute magnitudes of galaxies are computed in fixed bandpasses,
shifted to $z=0.1$, using Galactic reddening corrections 
(Schlegel, Finkbeiner, \& Davis 1998)
and $K$-corrections as described by Blanton et al. (2003). Therefore, all galaxies 
at $z=0.1$ have a $K$-correction of $-2.5{\rm log}(1+0.1)$, 
independent of their spectral energy distribution. We then apply the mean
luminosity evolution correction given by Tegmark et al. (2004), $E(z)=1.6(z-0.1)$.
The comoving distance limits of our volume-limited sample are 74.6 and 129.8 $h^{-1}$Mpc
if we adopt a flat $\Lambda$CDM cosmology with density parameters $\Omega_m=0.27$ 
and $\Omega_{\Lambda}=0.73$.
The galaxies in the volume-limited sample are divided into early (E and S0)
and late (S and Irr) morphological types based on the location of galaxies
in the $u-r$ color, $g-i$ color gradient, and concentration index space
(Park \& Choi 2005).

There are 14,032 late-type galaxies in the data set and, among them,
$\sim$8,700 galaxies have matching data in the 2 Micron All-Sky Survey (2MASS) 
Extended Source Catalog.
If the distance between the center of a SDSS galaxy and that of a 2MASS galaxy is
less than the semimajor axis in the SDSS $i$-band, 
we consider they are identical.
When there are multiple matches, we simply discard the data.
We also remove data when the absolute value of the color gradient parameter
($\Delta(g-i)$; see Park \& Choi 2005) 
is larger than 0.7, the $u-r$ color is larger that 4 or less than 0, or the 
H$_{\alpha}$ line width is less than 0 or larger than 200{\AA}.
About 700 galaxies have been discarded from this procedure.

\subsection{Parameters}

{\bf 1) $M_r$, $M_u$, and $u-r$:}  We use $r$-band absolute Petrosian magnitudes
and $u-r$ model color {}from the SDSS data.  
We obtain $u$-band absolute magnitude $M_u$ from the relation
\begin{equation}
   M_u = M_r + (u-r).
\end{equation}
Since the uncertainties in the extinction-corrections are large, we ignore 
the differences between the Petrosian $u$-band absolute magnitude and that derived from 
Equation (2).

{\bf 2) $c$:} The concentration index is defined by
   $R_{90}/R_{50}$ where $R_{50}$ and $R_{90}$ are the semimajor axis lengths
of ellipses containing 50\% and 90\% of the Petrosian flux in the SDSS $i$-band
image, respectively. It is corrected for the seeing effects by using the method
described in Park \& Choi (2005).  It is basically a numerical inverse mapping
from the image convolved with the PSF to the intrinsic image having
a Sersic profile and a fixed inclination.
The concentration index used here is the inverse of the one used by Park \& Choi.

{\bf 3) $a/b$:} The isophotal axis ratio $a/b$ is from the SDSS $i$-band image.
Here, $a$ is the major axis length and $b$ the minor axis length,
corrected for the seeing effects.
  We choose the isophotal axis ratio because the isophotal position angles
correspond most accurately to the true orientation of the major axis, which
is true for barred galaxies in particular. Here we assume that outside the central
region, the disk of a late-type galaxy can be approximated as a circular disk,
which thus appears as an ellipse in projection. If the disk of
   late-type galaxies has an intrinsic non-circularity, our estimation
   of the $a/b$ ratio will have some error due to our assumption.

{\bf 4) $M_K$:} We use $K_s$-band magnitude (more precisely, 
magnitude within the 20th mag arcsec$^{-2}$ elliptical isophote
set in $K_s$ band)
{}from the 2 Micron All-Sky Survey (2MASS) data.
We also apply $K$-corrections and Galactic extinction corrections for the 2MASS data
as described by Masters, Giovanelli, \& Haynes (2003).
As in the SDSS case, the $K$-corrections are made for a fixed redshift of $z=0.1$.
Most matched galaxies in our sample have $M_K$ brighter than $-20$.

\begin{figure}[t]
\plotone{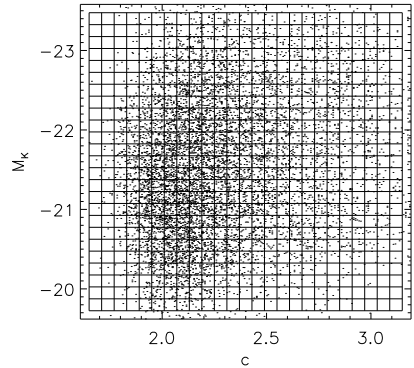}  
\caption{
Distribution of the galaxies in our volume-limited sample in the
$K_s$-band absolute magnitude $M_K$ and the concentration index
$c=R_{90}/R_{50}$ space divided into $25\times25$ cells.}
\label{fig_ckplane} 
\end{figure}

\subsection{Method}

We adopt the popular parameterization of the effects of internal extinction:
\begin{equation}
   A_{\lambda} = \gamma_{\lambda} \log_{10}(a/b),
\end{equation}
where $A_{\lambda}$ is the amount of extinction.
In this parameterization, $\gamma_{\lambda}$ is the slope
of a scatter plot drawn on the $\log_{10}(a/b)$ - $A_{\lambda}$ plane.
We will see in section 4 that this is actually a very reasonable model in the case of the
SDSS galaxies even though some recent studies reported that better models
can be found (Masters et al. 2003; Rocha et al. 2008; Unterborn \& Ryden 2008).

Our goal is to find the values of $\gamma_r$ and $\gamma_u$.
To find $\gamma_{r}$ (or $\gamma_u$), we first plot $r-K_s$ (or $u-K_s$)
against $\log_{10}(a/b)$. 
Then, the slopes of the scatter plot are 
\begin{eqnarray}
  \gamma_{r-K} &\approx &\gamma_r - \gamma_K, \mbox{~~~~~ and} \nonumber \\
  \gamma_{u-K} &\approx &\gamma_u - \gamma_K.
\end{eqnarray}
When no strong extinction is present in $K_s$ band, we can write
\begin{eqnarray}
  \gamma_{r-K} &\approx &\gamma_r,  \mbox{~~~~~ and} \nonumber \\
  \gamma_{u-K} &\approx &\gamma_u.
\end{eqnarray}
In this paper, we assume that $K_s$-band magnitude is almost
free of inclination effects. 
Masters et al.~(2003) analyzed galaxies in the 2MASS Extended
Source Catalog and concluded that the internal extinction is indeed
small in $K_s$ band. 
Their results are consistent with the earlier result that $\gamma_K \sim 0.22$
(Tully et al.~1998).

\begin{figure*}[t]
\begin{center}
\plotone{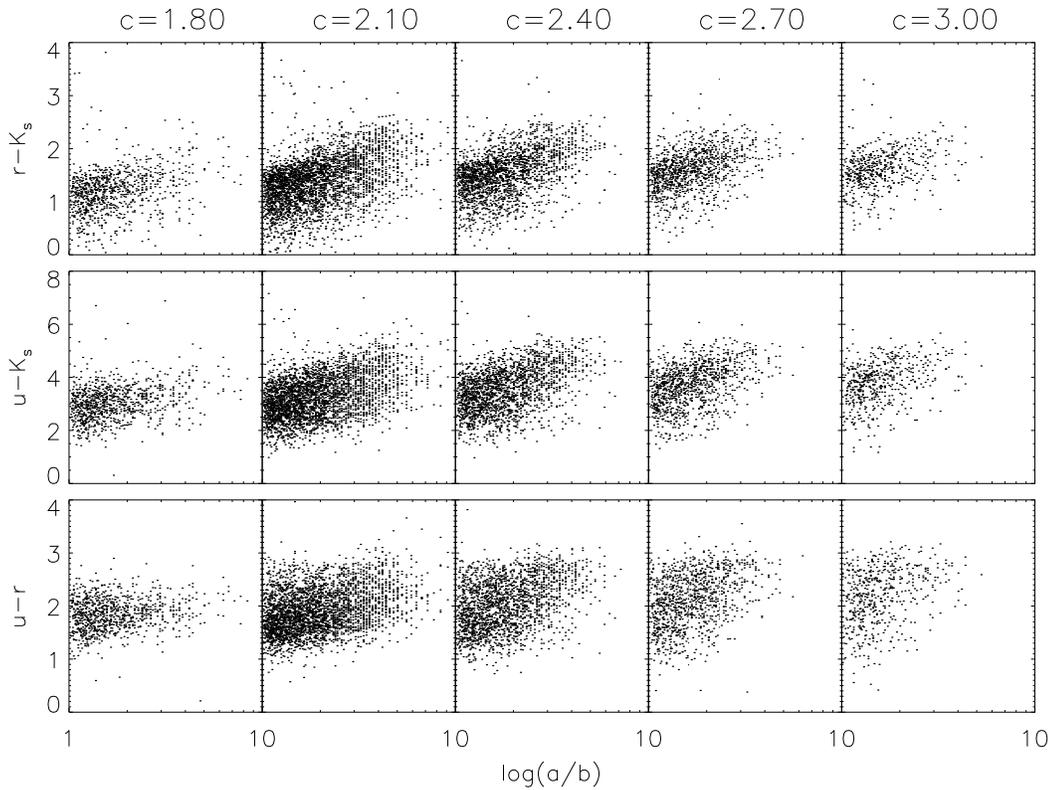}
\caption{ Dependence of colors on the concentration index $c$ ($\equiv R_{90}/R_{50}$).
  When $c$ is small, the colors only weakly depend on
  the axial ratio $a/b$. The axial ratio
  is derived from the SDSS $i$-band image. }
\label{fig_2}
\end{center}
\end{figure*}
\begin{figure*}[t]
\epsscale{0.32}
\plotone{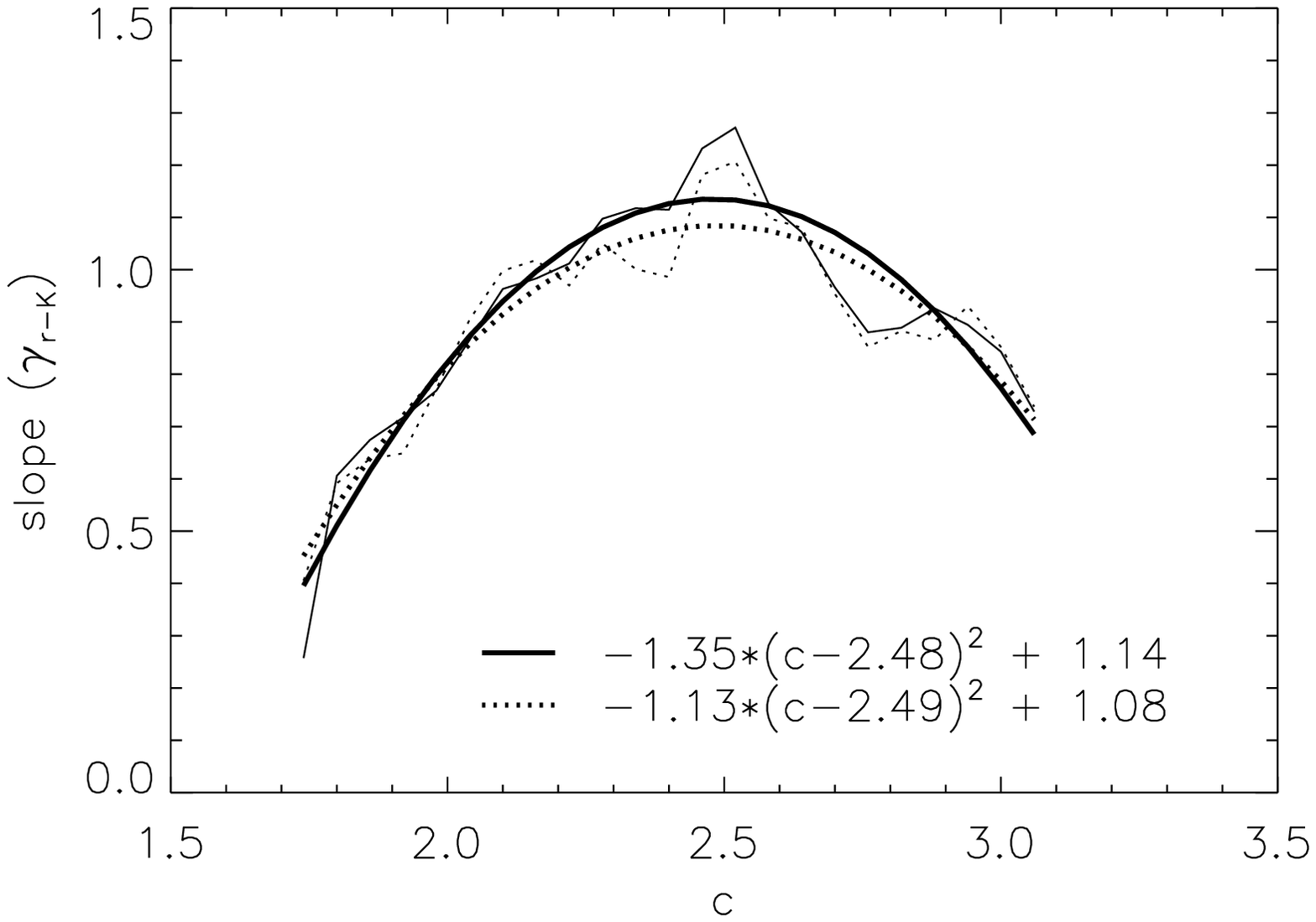}
\epsscale{0.32}
\plotone{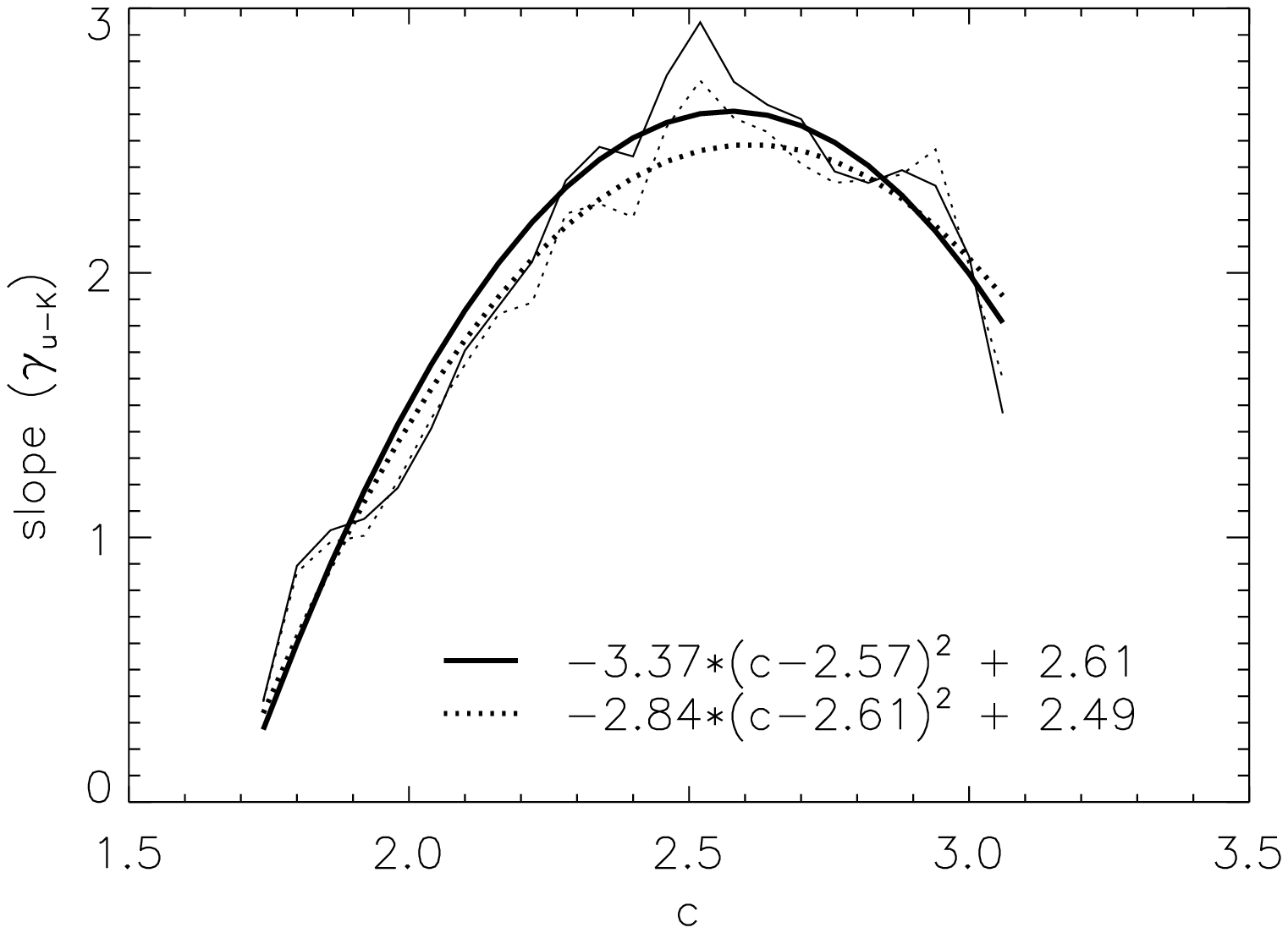}
\epsscale{0.32}
\plotone{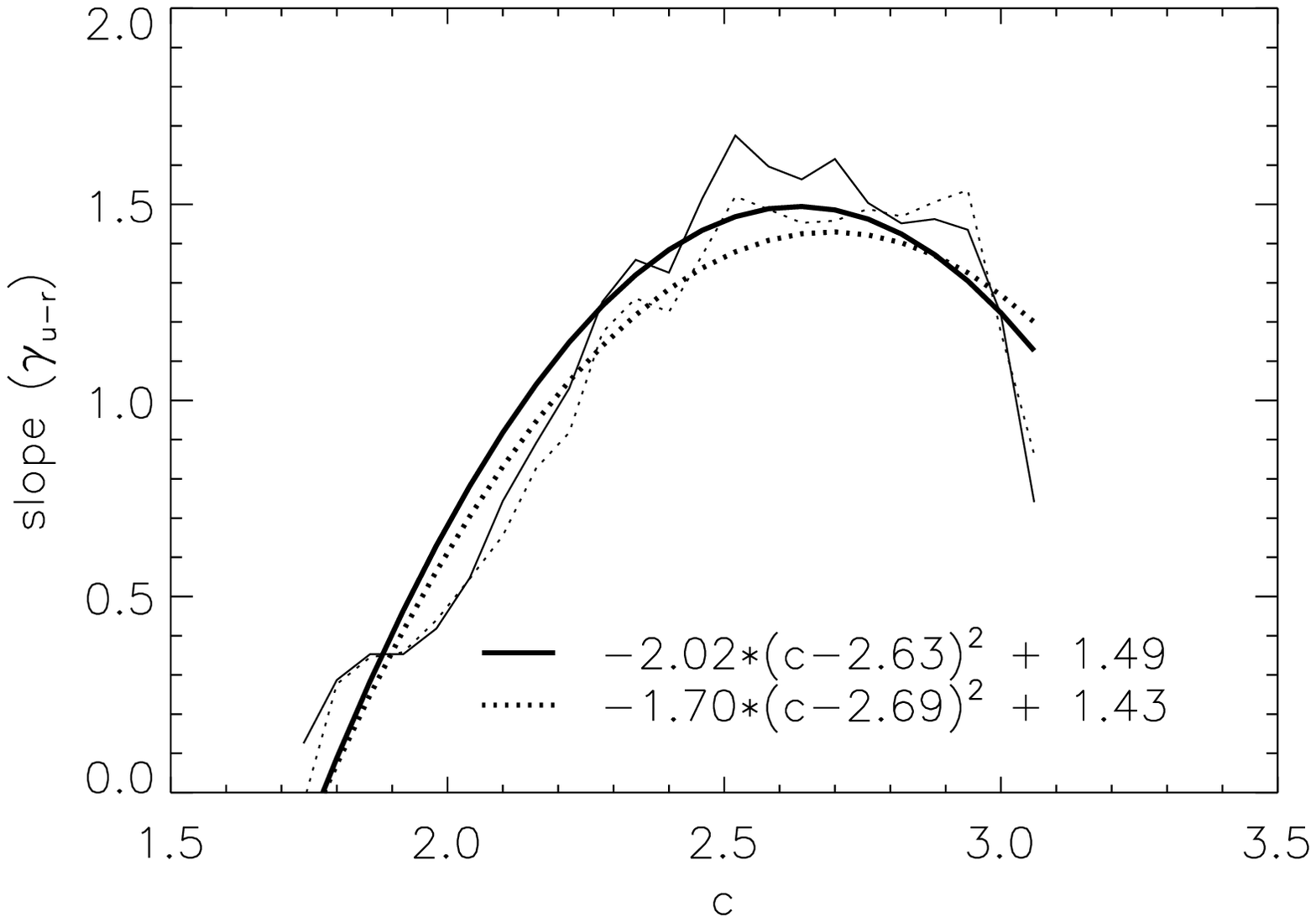}
\caption{ Dependence of the slope of the extinction law on $c$ ($\equiv R_{90}/R_{50}$).
 Solid lines are the slopes of the linear fit to
 the most probable values in $\log(a/b)$ bins between $\log(a/b)=0$ and 0.5.
 Dotted lines are the slopes of the linear fit to
 the median.
 ({\it Left panel}) $\gamma_{r-K}$.
 ({\it Middle panel}) $\gamma_{u-K}$.
 ({\it Right panel}) $\gamma_{u-r}$.
}
\label{fig_slope_c}
\end{figure*}

As shown by earlier studies, $\gamma_{\lambda}$ depends on luminosity
(Giovanelli et al.~1995; Tully et al.~1998) 
and/or galaxy type (de Vaucouleurs et al.~1991; Han 1992).
In this study, our primary concern is the dependence of $\gamma_{\lambda}$ on
$K_s$-band luminosity and the concentration index $c$.
In Figure~\ref{fig_ckplane}, we plot the galaxies on the $c-M_K$ plane.
On the plot, we show $25\times 25$ cells.
The coordinates of the cell centers, $(c_m, K_n)$, are given by
\begin{eqnarray}
    c_m &=& 1.68 + 0.06m, \mbox{~~~m= 0, 1,..., 24,~~~ and}  \\
    K_{n} &=& -23.4+0.15n, \mbox{~~~n= 0, 1,..., 24.} 
\end{eqnarray}
Note that $c_2=1.8$, $c_{22}=3.0$, $K_2=-23.1$, and $K_{22}=-20.1$.

When we investigate the dependence of $\gamma_{\lambda}$ on $M_K$ and $c$,
we may simply use galaxies in each cell in Figure~\ref{fig_ckplane}.
However, some cells in Figure~\ref{fig_ckplane} are not sufficiently populated.
Therefore, for smooth results, 
we use galaxies in $5\times 5$ cells 
for scatter plots. 
   More precisely, we use galaxies with 
   $c_m-2.5(\Delta c) \leq c \leq c_m+2.5(\Delta c)$ and
   $K_n-2.5(\Delta K) \leq M_K \leq K_n+2.5(\Delta K)$, where
   $\Delta c=0.06$ and $\Delta K=0.15$.
We allow for a similar overlap
when we study $M_K$- or $c$-dependence separately.

To find the slopes, 
we divide $\log(a/b)$ axis into 20 bins between 0 and 1.
Then we find a representative value for each bin.
We try two methods to obtain the representative value:
$$
\begin{array}{l}
\mbox{1) the median} \\
\mbox{2) the most probable value.}
\end{array}
$$
To fine the most probable value in each $\log(a/b)$ bin, we use a smoothing function
\begin{eqnarray}
\phi(x,y)=
 \left\{
  \begin{array}{ll}
    \exp{(-(y-y_i)^2/2\sigma_y^2)} & \mbox{ if $|x-x_i|$ $\leq$ 0.05}
\\
    0 & \mbox{    otherwise,}
  \end{array} \right.
\end{eqnarray}
where $x=\log_{10}(a/b)$, $x_i$ ($y_i$) is the $x$ ($y$) value of the $i^{th}$ galaxy,
and $\sigma_y$ is the standard deviation.  
After applying the smoothing function, we find the maximum value
of the smoothed distribution in each bin.
After finding representative values in bins of $\log(a/b)$,
we perform the linear fit to the representative values between
$\log(a/b)=0$ and 0.5.


\begin{figure*}[t]
\begin{center}
\epsscale{1.0}
\plotone{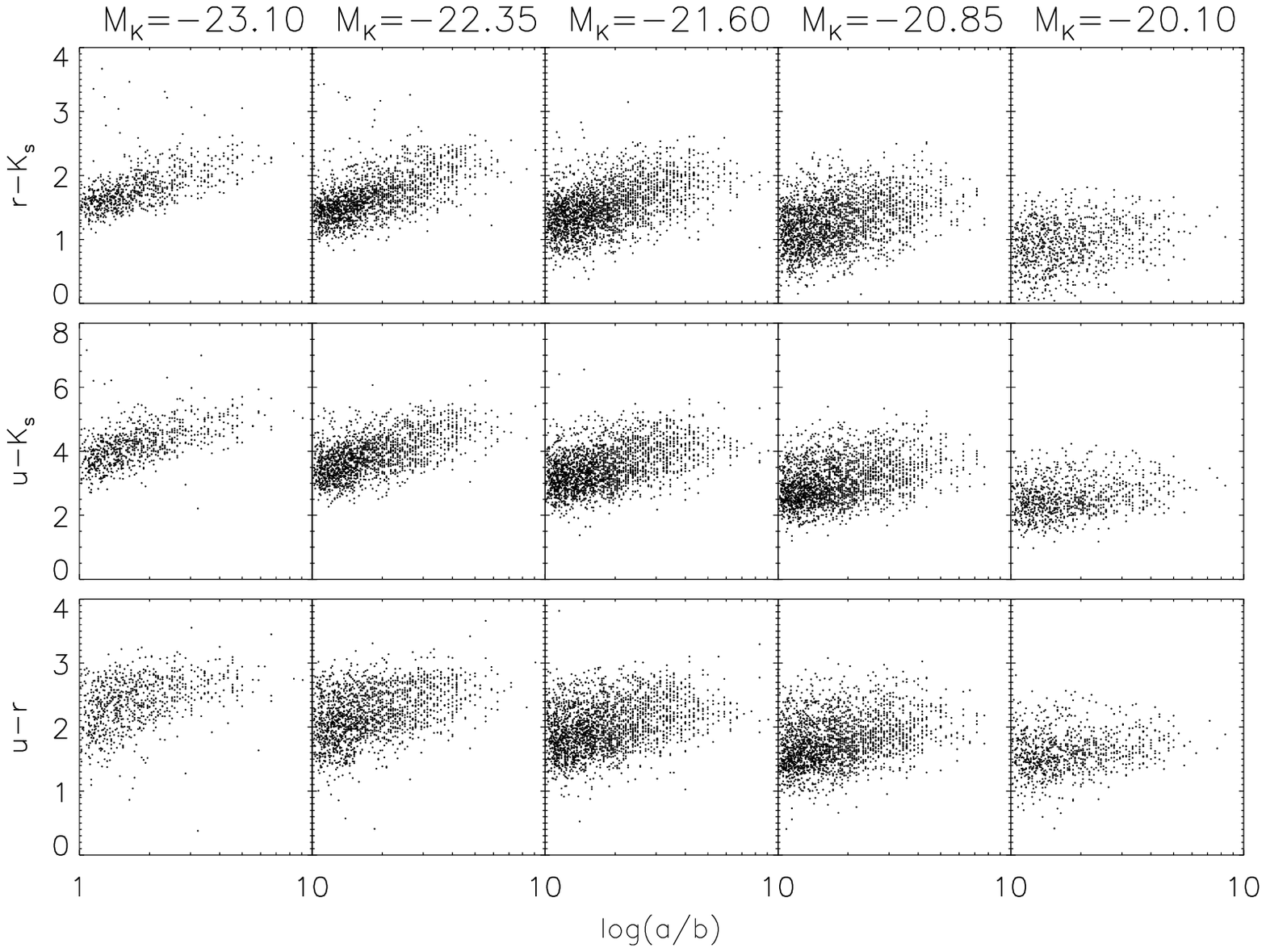}
\caption{ Dependence of colors on $M_K$.
 Faint galaxies in $K_s$ band exhibit shallower slopes
 in all 3 colors, while bright galaxies show steep slopes.}
\label{fig_4}
\end{center}
\end{figure*}
\begin{figure*}[t]
\epsscale{0.32}
\plotone{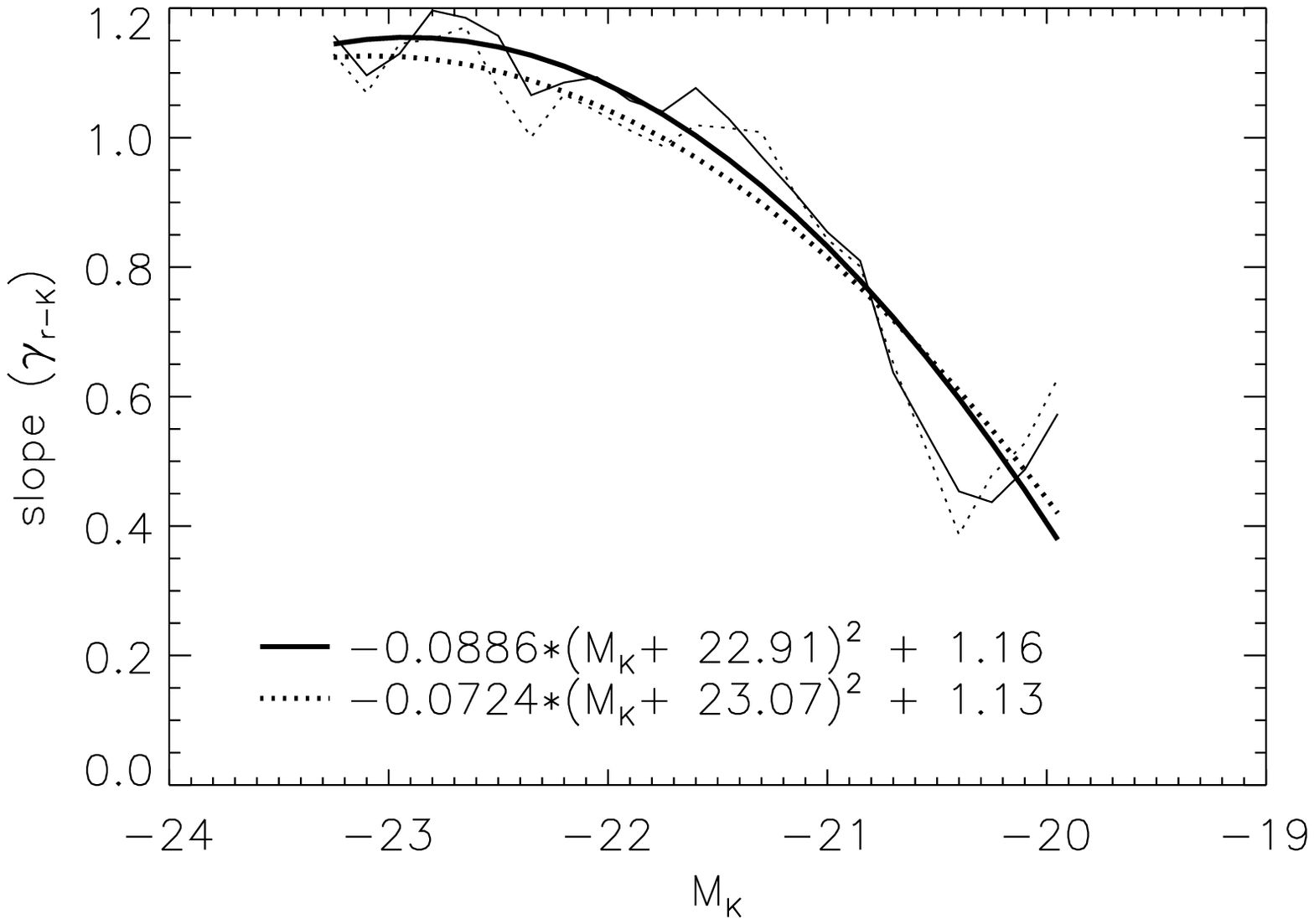}
\epsscale{0.32}
\plotone{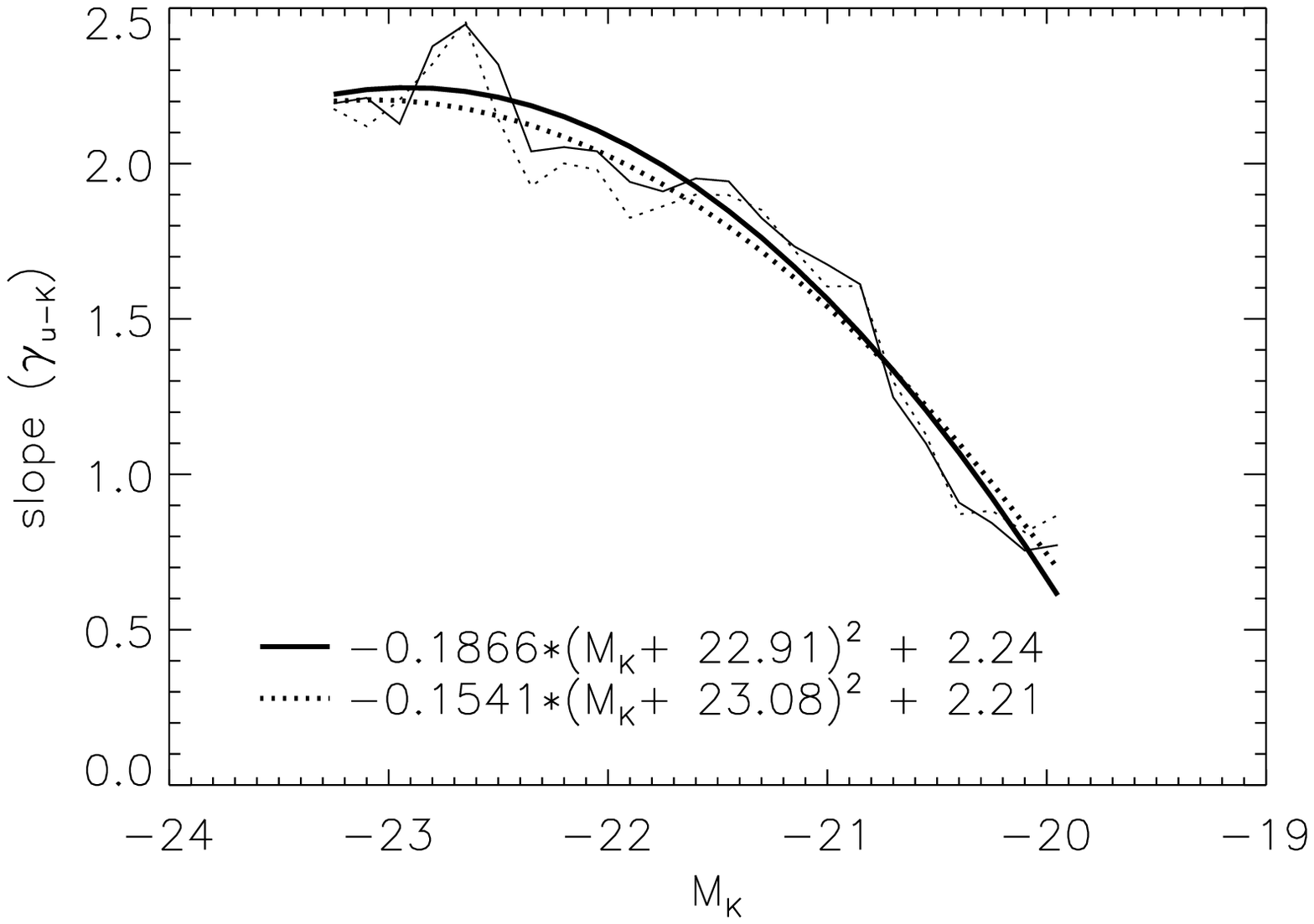}
\epsscale{0.32}
\plotone{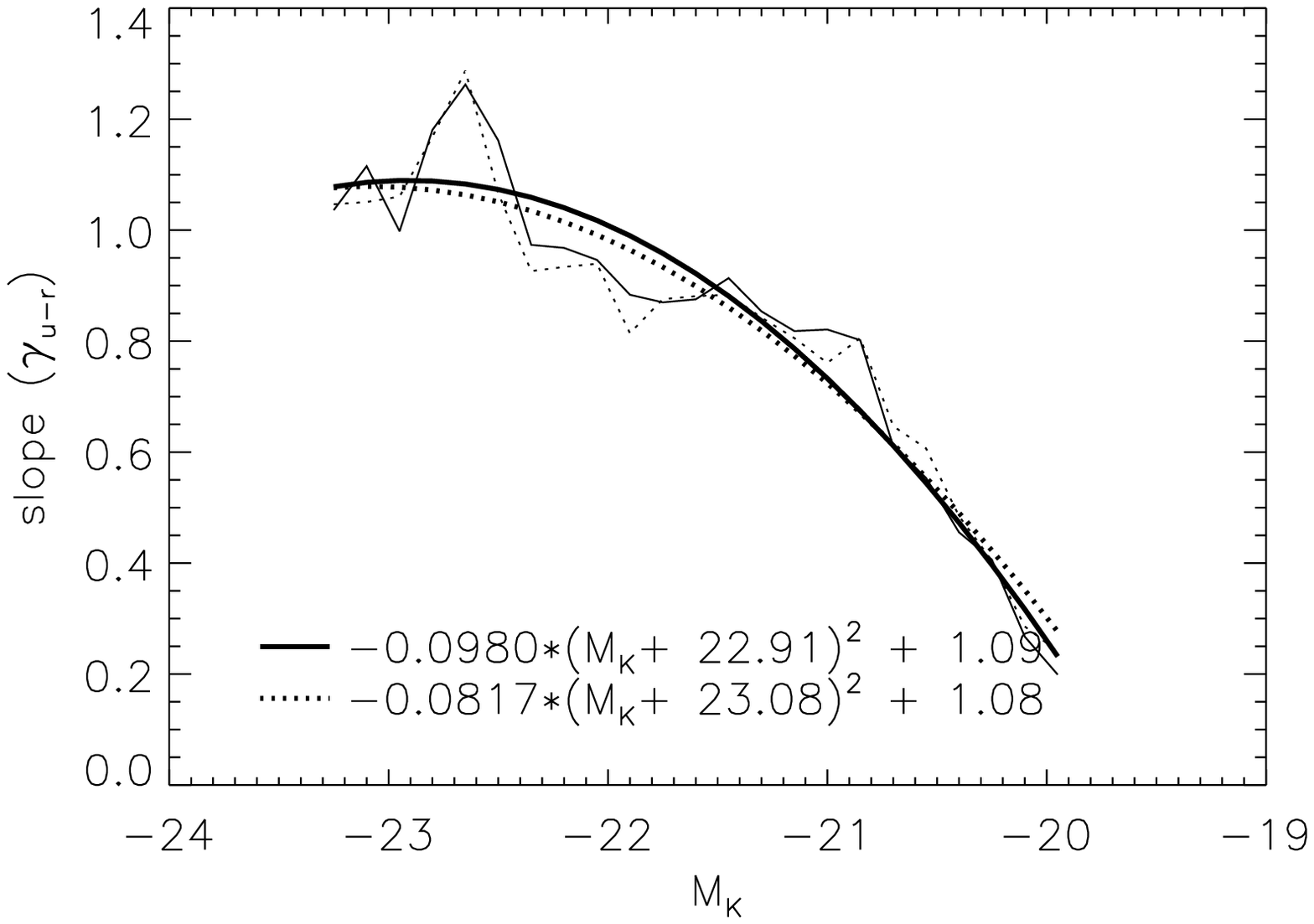}
\caption{ Dependence of the slope of the extinction law on the $K_s$-band absolute magnitude $M_K$. 
Solid lines are the slopes of the linear fit to
the most probable values in $\log(a/b)$ bins between $\log(a/b)=0$ and 0.5.
Dotted lines are the slopes of the linear fit to
the median.
({\it Left panel}) $\gamma_{r-K}$.
({\it Middle panel}) $\gamma_{u-K}$.
({\it Right panel}) $\gamma_{u-r}$.
}
\label{fig_slope_k}
\end{figure*}

\section{Dependence on $c$ or \MakeUppercase{$M_K$}}
\subsection{Dependence on $c$}
In this subsection, we consider only $c$-dependence.
We plot $r-K_s$, $u-K_s$, and $u-r$ colors against $\log_{10}(a/b)$ in Figure~\ref{fig_2}.
As we mentioned earlier, when we draw the scatter plot for $c=c_m$, 
we use galaxies with $c_m-2.5(\Delta c) \leq c \leq c_m+2.5(\Delta c)$, where
$\Delta c=0.06$. Therefore, the plot for $c=1.80$ contains galaxies
with $1.65 \leq c \leq 1.95$, for example.

We can see that the slope, hence $\gamma_{\lambda}$,
 for $c=1.80$ is smaller than those for other $c$
values. 
Note that the slope for $c=1.80$ is very close to zero for $u-r$ color.

Figure~\ref{fig_slope_c} shows dependence of the slope on $c$.
All 3 colors show a common feature: the slope peaks
near $c\sim 2.5$.
This means that internal extinction is maximum for intermediate late-type galaxies,
and is smaller for early and late late-type galaxies. On the other hand,
$u-r$ and $u-K_s$ show stronger dependence on $c$
than $r-K_s$, telling that internal extinction is higher in shorter wavelength bands.
The quadratic equations on the plots are the fitting functions.
The solid curves are for the most probable values and the dotted curves for
the median. The RMS scatters of the measured slope from the fitting functions 
(for the most probable values) are
0.071, 0.170, and 0.154 for $r-K_s, u-K_s$, and $u-r$ colors, respectively.
We list the quadratic fits, $\gamma(c)$'s, in Table 1 and 2.



\begin{figure*}[p]
\begin{center}
\epsscale{1.0}
\plotone{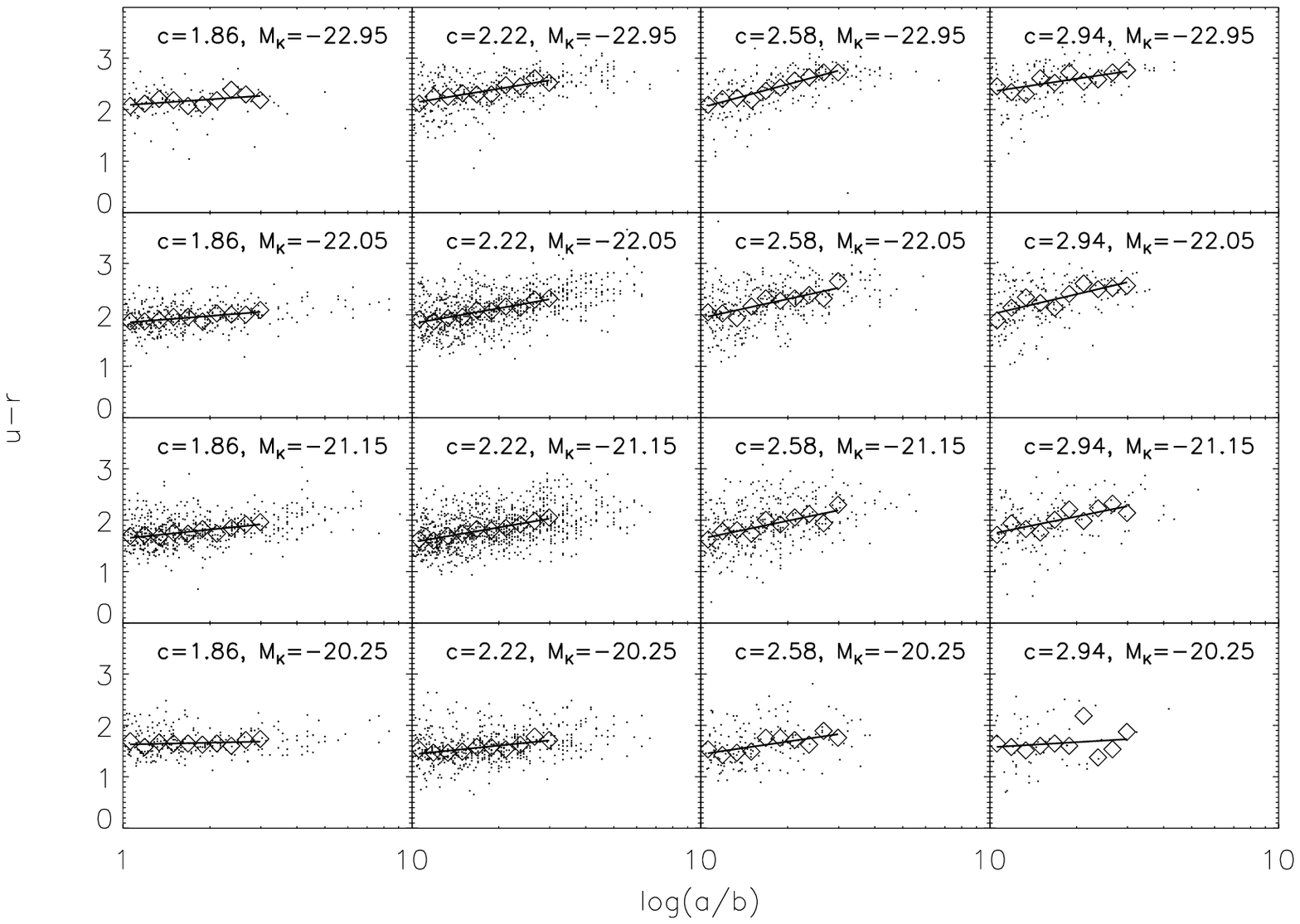}
\caption{ 
Scatter plots for $u-r$.  
Diamonds denote the most probable values in $\log(a/b)$ bins.
Lines are the results of the least square fit to the diamonds. 
   Only points with $\log_{10}(a/b) \le 0.5$ are used for fitting.
}
\label{fig_2d_ur}
\end{center}
\end{figure*}

\begin{figure*}[p]
\begin{center}
\includegraphics[angle=-90,width=0.63\textwidth]{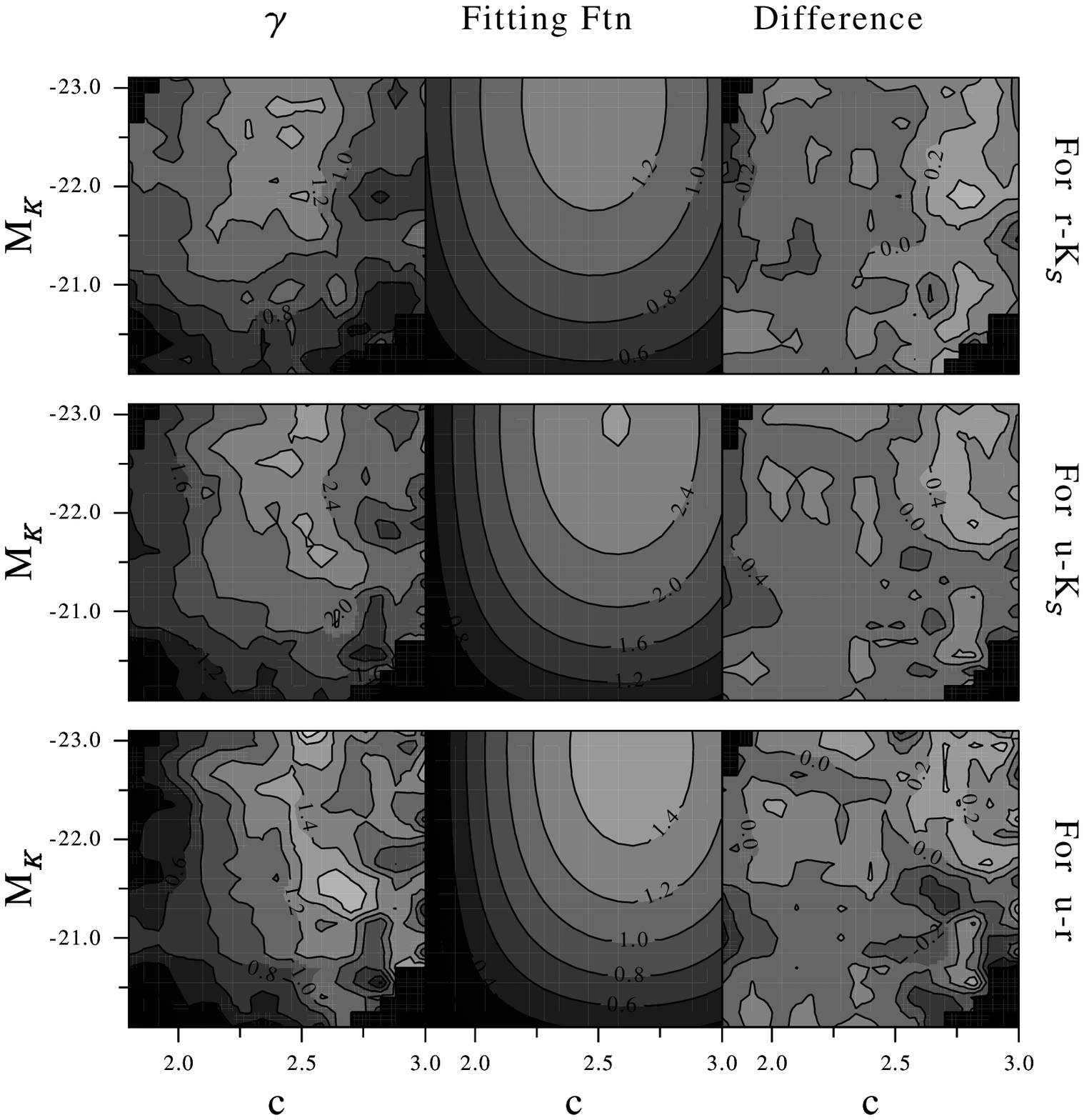} 
\caption{
The measured slopes of the extinction law (left panels), 
fitting functions (middle panels),
and the difference between the measured slopes 
and the fitting functions (right panels).
({\it Upper panels}) $r$-$K_s$ color.
({\it Middle panels}) $u$-$K_s$ color.
({\it Lower panels}) $u-r$ color.
The horizontal axis of each contour plot is the concentration index $c$ 
($\equiv R_{90}/R_{50}$)
and the vertical axis is the $K_s$-band absolute magnitude $M_K$.
The contours are drawn at intervals of 0.2 (upper panels), 0.4 (middle
panels), and 0.2 (lower panels), respectively.
} 
\label{fig_9conto}
\end{center}
\end{figure*}

\subsection{Dependence on \MakeUppercase{$M_K$}} 
In this subsection, we consider $M_K$-dependence of internal extinction.
We plot $r-K_s$, $u-K_s$, and $u-r$ colors against $\log_{10}(a/b)$ in Figure~\ref{fig_4}.
When we draw the scatter plot for $M_K=K_n$, 
we use galaxies with $K_n-2.5(\Delta K) \leq M_K \leq K_n+2.5(\Delta K)$, where
$\Delta K=0.15$. For example, the plot for $M_K=-23.1$ contains galaxies
with $-23.475 \leq M_K \leq -22.725$.

We can clearly see that the slope, hence $\gamma_{\lambda}$,
 for $M_K=-20.1$ is smaller than those for other $M_K$
values. 
Note that the slope for $M_K=-20.1$ is very close to zero for all 3 colors.

Figure~\ref{fig_slope_k} shows dependence of the slope on $M_K$.
All 3 colors have a common feature: the slope is higher
when galaxies are brighter in $K_s$ band.
Due to lack of data points, the slope is not clear
for $M_K \lesssim -23.2$.
Therefore, one should be careful when using the fitting functions shown on the plots 
for galaxies with $M_K\lesssim-23.2$.
The RMS scatters of the measured slope from the fitting functions 
(for the most probable values) are
0.071, 0.110, and 0.075 for $r-K_s, u-K_s$, and $u-r$ colors, respectively.
We list the quadratic fits, $\gamma(M_K)$'s, in Table 1 and 2.

\section{Dependence on $c$ and \MakeUppercase{$M_K$}}
We now study dependence of $r-K_s, u-K_s,$ and $u-r$ colors on
the concentration index and $K_s$-band absolute magnitude.
In Figure~\ref{fig_2d_ur} we plot $u-r$ against $\log_{10}(a/b)$
as a function of $c$ and $M_K$.
The diamonds are the most probable values in bins of $\log(a/b)$.
The lines are the least square fits to the most probable values.
It can be seen that the reddening is fit well
by our extinction model linear in ${\rm log}_{10}(a/b)$ (i.e. Equ. 3), 
and that the slope depends on both $c$ and $M_K$.
Plots for $r-K_s$ and $u-K_s$ show behaviors similar to the $u-r$ case. 
We also obtained
the fits to the median, which are qualitatively similar to the most probable case.

In the top panels of Figure~\ref{fig_9conto} we present contour plots of
the measured slopes of $r-K_s$ color, $\gamma_{r-K}$.
The upper-left panel is for $\gamma_{r-K}$ based on the most probable 
values. 
The function
\begin{equation}
     \gamma_{r-K}(c,M_K) =1.02 \gamma_{r-K}(c) \gamma_{r-K}(M_K),
\label{eq_9}
\end{equation}
shown in the upper-middle panel,
fits well the slopes based on the most probable values (see Table 1).
Since we do not have enough bright galaxies in $K_s$ band,
we do not have a reliable fitting formula for $M_K<-23.25$.
Although our fitting formula suggests that the slope
declines when $M_K$ becomes less than $\sim-22.9$, 
the true behavior of the slope may be different.
Therefore, instead of using the fitting formula in Equation (\ref{eq_9}),
one may use 
\begin{equation}
  \gamma_{r-K}(c,M_K) =1.02 \times 1.14  \gamma_{r-K}(c) 
\label{eq_rk_max}
\end{equation}
for $M_K<-22.9$.
The upper-right panel of Figure \ref{fig_9conto} shows the difference between the measured slope
and the fitting function. The RMS value of the difference is 0.174 while the peak
slope is about 1.4.
See Table 2 for a fitting function based on the median.

Similarly, we present contours of $\gamma_{u-K}$ and 
$\gamma_{u-r}$ in the middle and bottom panels of Figure~\ref{fig_9conto}, respectively.
The fitting functions for the most probable values are
\begin{eqnarray}
   \gamma_{u-K}(c,M_K)=0.48\gamma_{u-K}(c) \gamma_{u-K}(M_K), \\
   \gamma_{u-r}(c,M_K)=0.94\gamma_{u-r}(c) \gamma_{u-r}(M_K) 
\end{eqnarray}
(see Table 1).
Again the fitting functions are uncertain for $M_K\lesssim -23.2$.
The RMS differences, shown in the middle-right and
lower-right panels of Figures \ref{fig_9conto},
are 0.354 and 0.234, respectively.
See Table 2 for fitting functions based on the median.

\begin{figure*}[p]
\begin{center}
\plotone{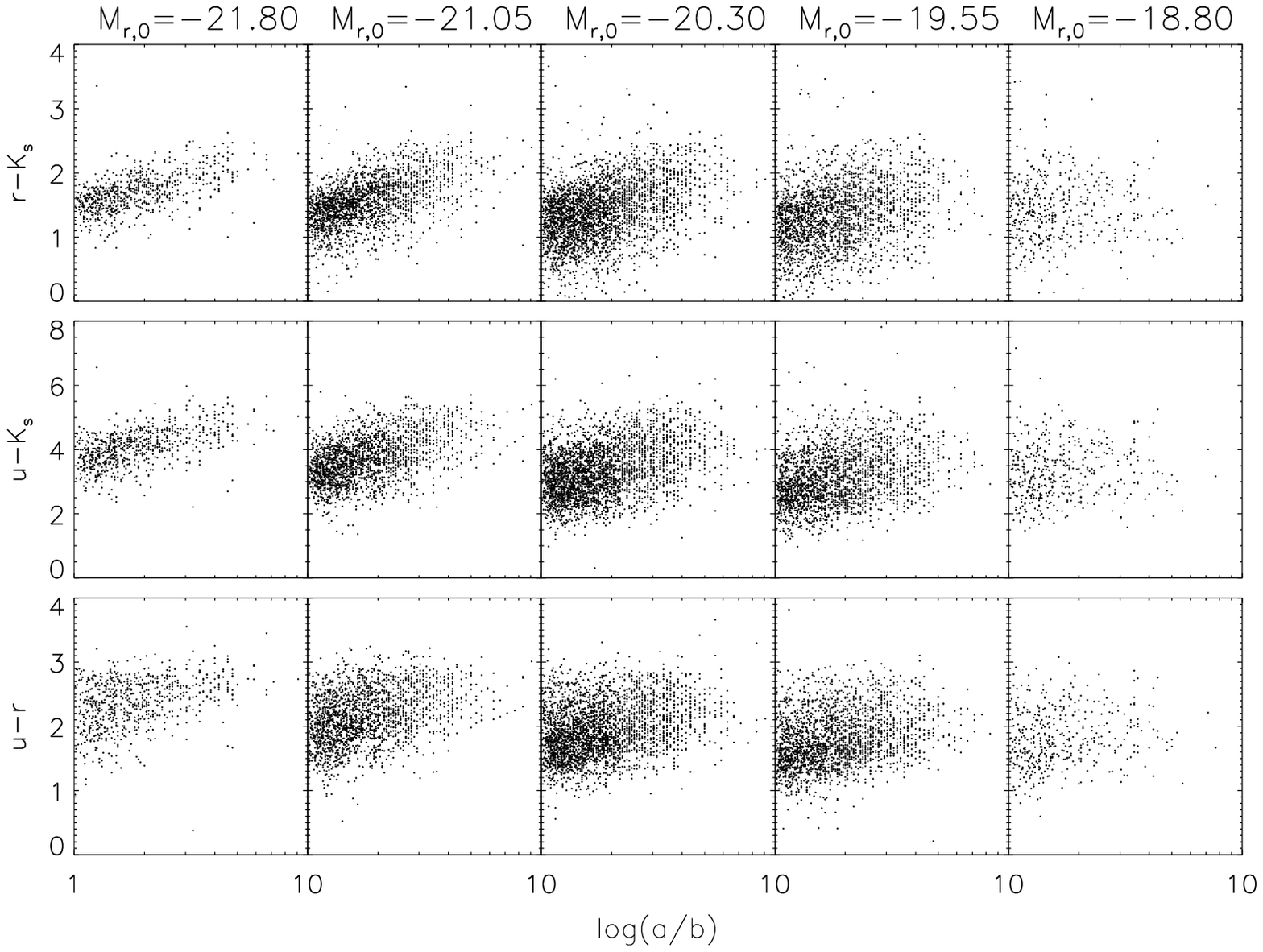}
\caption{ Dependence of colors on the intrinsic (i.e. face-on) 
$r$-band absolute magnitude $M_{r,0}$.
  We use $\gamma_R \sim \gamma_{r-K} \sim \gamma_{r-K}(c,M_k)$ (see Table 1)
  to derive $M_{r,0}$. 
  When $M_{r,0} \lesssim -19$, internal extinction is very small.}
\label{fig_1d_r}
\end{center}
\end{figure*}
\begin{figure*}[p]
\epsscale{0.32}
\plotone{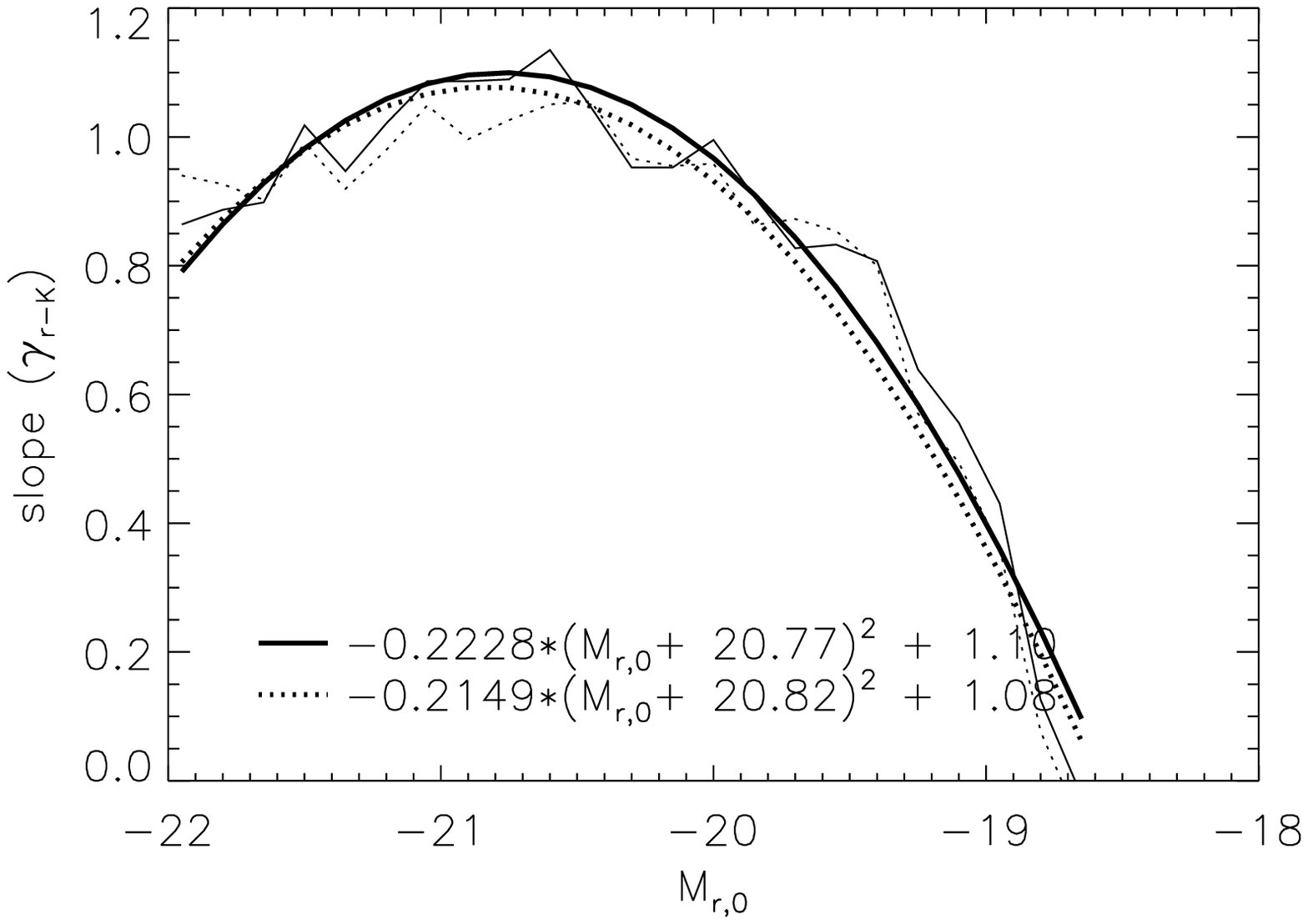}
\epsscale{0.32}
\plotone{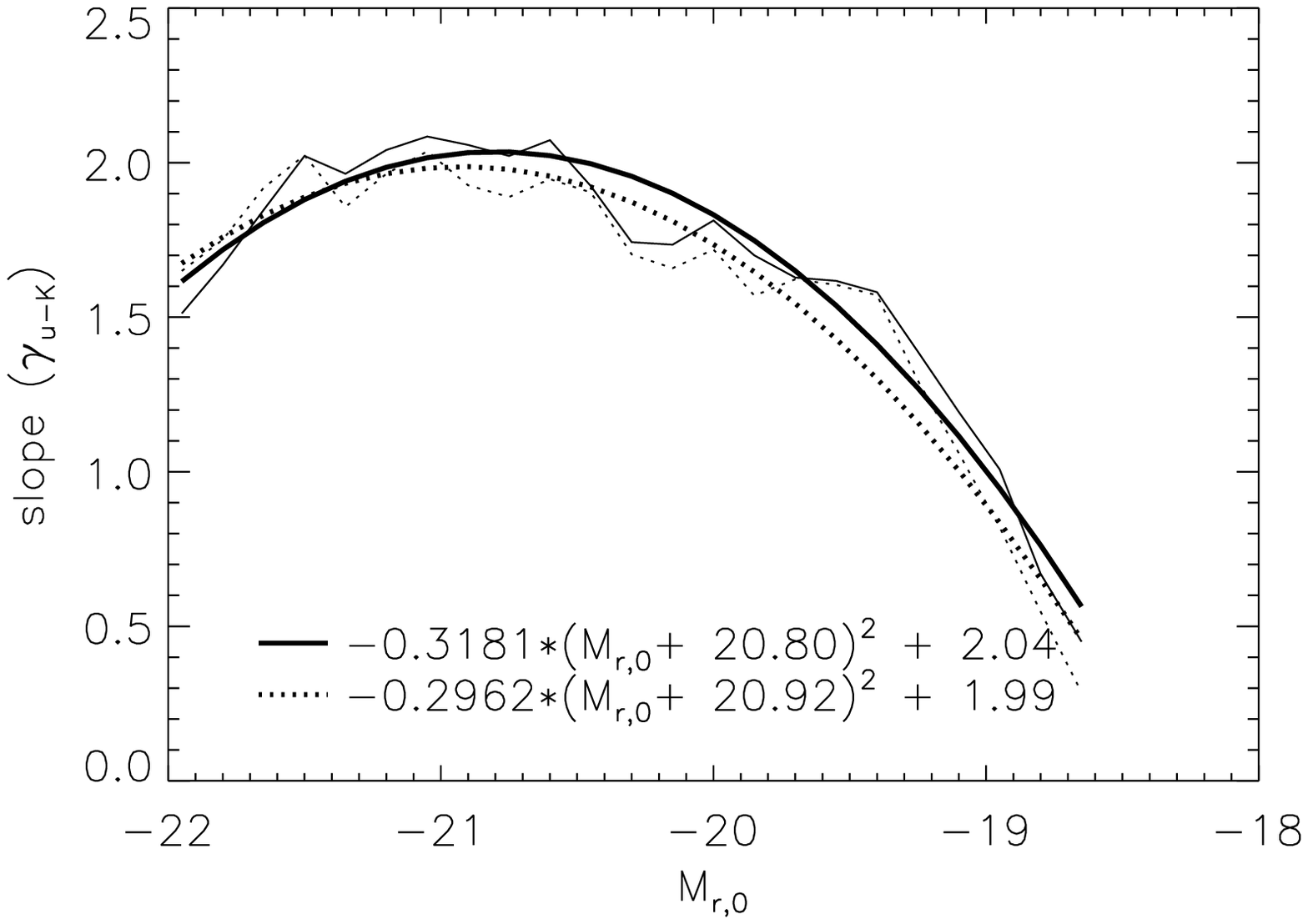}
\epsscale{0.32}
\plotone{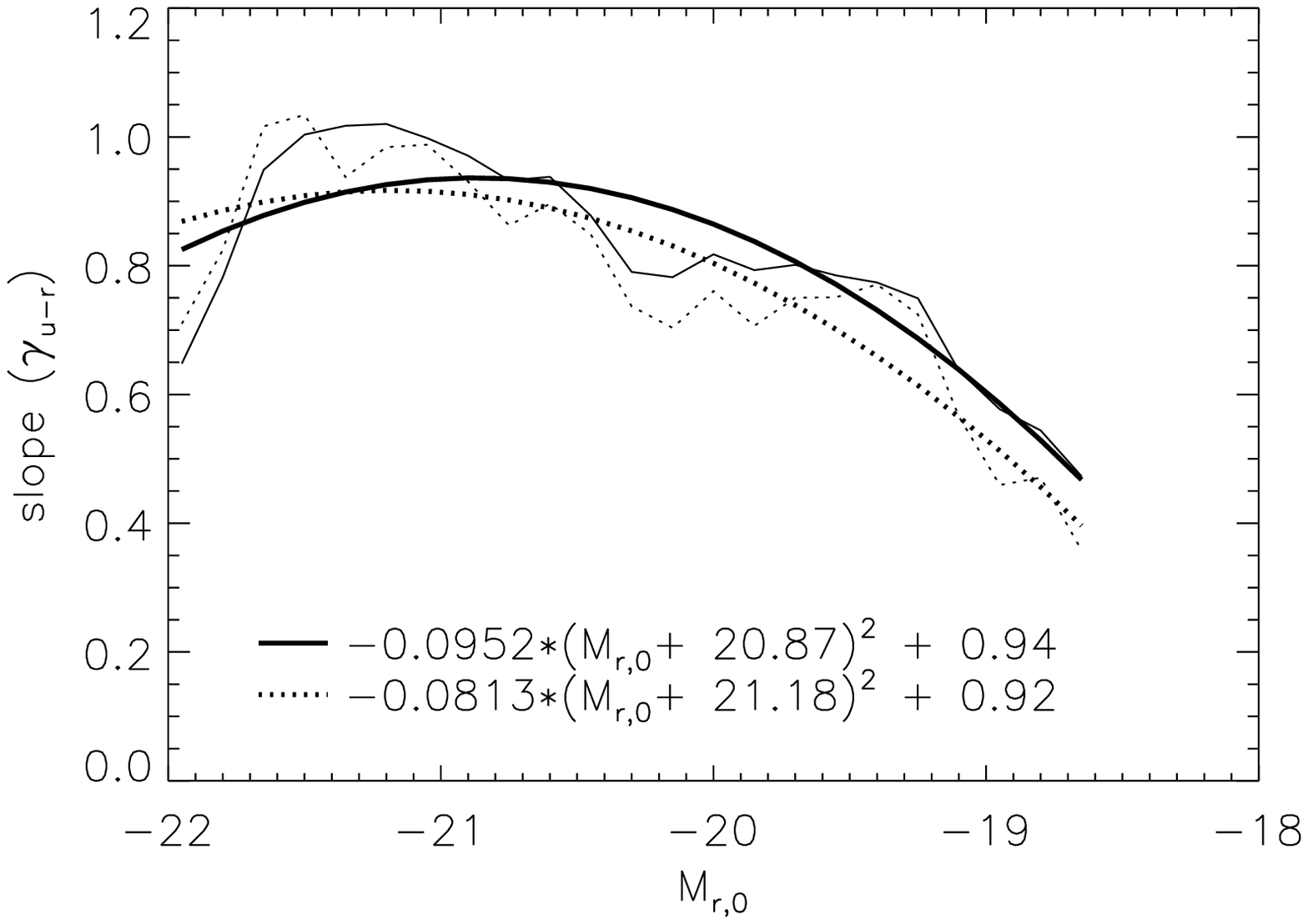}
\caption{ Dependence of $\gamma$ on the intrinsic (i.e. face-on) $r$-band absolute magnitude $M_{r,0}$.
Solid lines are the slopes of the linear fit to
the most probable values in $\log(a/b)$ bins between $\log(a/b)=0$ and 0.5.
Dotted lines are the slopes of the linear fit to
the median.
The RMS scatters of the measured slope from the fitting functions 
(for the most probable values) are
0.064, 0.095, and 0.070 for $r-K_s, u-K_s$, and $u-r$ colors, respectively.
({\it Left panel}) $\gamma_{r-K}$.
({\it Middle panel}) $\gamma_{u-K}$.
({\it Right panel}) $\gamma_{u-r}$.
}
\label{fig_fitR}
\end{figure*}

\begin{figure*}[p]
\includegraphics[angle=-90,width=0.90\textwidth]{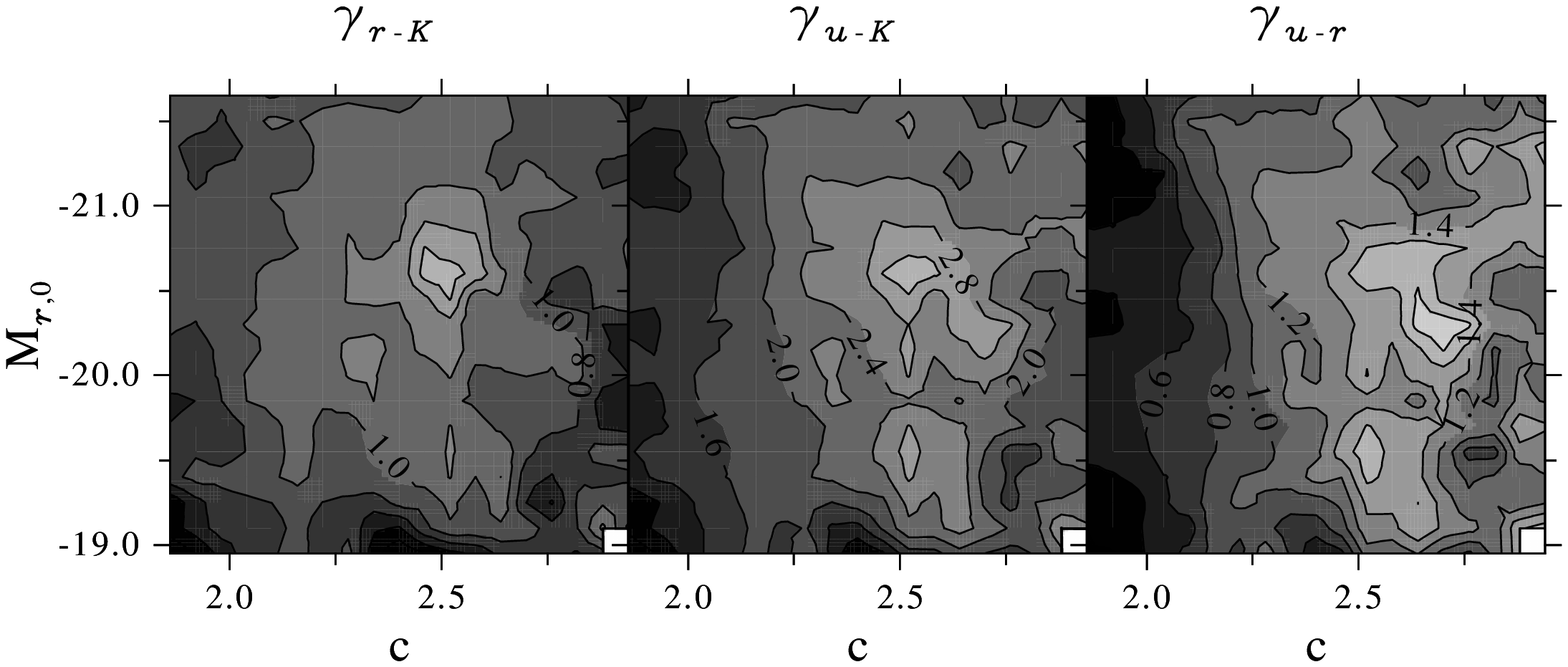} 
\caption{
Dependence of the slope of the extinction law for the $r-K$, $u-K$, and $u-r$ colors
on the $r$-band absolute magnitude $M_{r,0}$ and the concentration $c$.
     The slope $\gamma$ is based on the most probable values.
   X-axis is the concentration index ($=R_{90}/R_{50}$).
The contour intervals for $r$-$K_s$, $u$-$K_s$, and $u-r$ are 0.2, 0.4, and 0.2,
respectively.
} 
\label{fig_2d_cr}
\end{figure*}

\section{Dependence on \MakeLowercase{r} magnitude}

When $K_s$-magnitude is not available, we cannot use the fitting functions
derived in the previous section.
In this section we present a method that utilizes $r$-magnitude, instead of $K_s$-magnitude.
We consider $\gamma$'s derived from  the most probable values.

\subsection{Dependence on $M_{r,0}$}
We obtain intrinsic (or face-on) $r$-band absolute
magnitude, $M_{r,0}$, from
\begin{equation}
   M_{r,0}=M_{r,obs} - \gamma_{R}(c,M_K) \log_{10}(a/b),
\end{equation}
where $M_{r,obs}$ is the $r$-band absolute magnitude before inclination corrections and
$\gamma_r(c,M_K) \sim \gamma_{r-K}(c,M_K)$ is given in Table 1.
Note that $\gamma_{r-K}(c,M_K)\geq 0$.
The RMS differences between the measured slopes and the fitting functions given in the last
column of Table 1,
are
0.208, 0.390, and 0.233 for $r-K_s, u-K_s$, and $u-r$ colors, respectively.
After obtaining $M_{r,0}$ we investigate how internal extinction depends
on the intrinsic $r$-band luminosity.

Scatter plots in Figure~\ref{fig_1d_r} clearly shows 
that the slopes of the extinction in $r-K_s, u-K_s$, and $u-r$ colors
 depend on the $r$-band absolute magnitude $M_{r,0}$.
The slopes of the scatter plots are very small when $M_{r,0} = -18.80$, which
means there is very little extinction in galaxies much fainter than the $M_*$ galaxies. 
(But note that late-type galaxies tend to be irregulars
at fainter magnitudes and that irregulars have patchy distribution of dust
which can cause a weaker dependence of extinction on inclination.)

Figure~\ref{fig_2d_cr} show the behavior of the slopes on $c-M_{r,0}$ plane.
It is interesting that there is a well-defined peak near 
$(c,M_{r,0})\sim (2.5, -20.8)$.
In the $r$-band, the internal extinction is maximum at the absolute magnitude
very close to the characteristic magnitude $M_*$ (see Table 2 of Choi et al. (2007)
for $M_*$ measurements for the late-type SDSS galaxies. Note that $h=0.75$ is used
in the present work.)
The contour spacing for $\gamma_{r-K}$ (left panel) is 0.2 and
the value of $\gamma_{r-K}$  at the
center of the peak is greater than $\sim 1.6$.
We can also see a similar peak for $\gamma_{u-K}$ (middle panel).
However, the location of the peak for $\gamma_{u-r}$ (right panel) is
somewhat different.

Figure~\ref{fig_fitR} shows the dependence of the slope on $M_{r,0}$.
The fitting functions are also shown on the plots.

\begin{figure*}[h!t]
\includegraphics[angle=-90,width=0.32\textwidth]{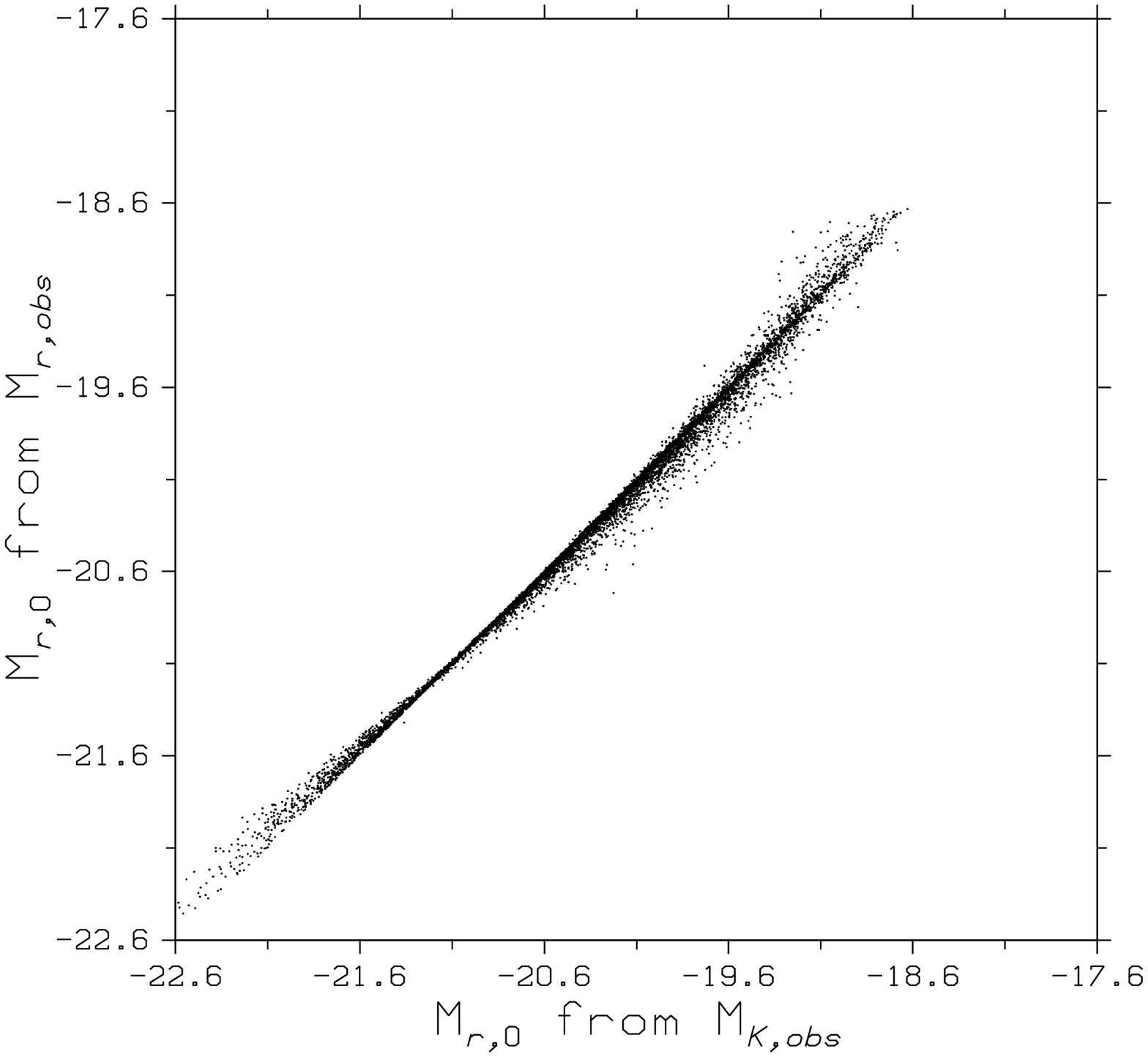}
\includegraphics[angle=-90,width=0.32\textwidth]{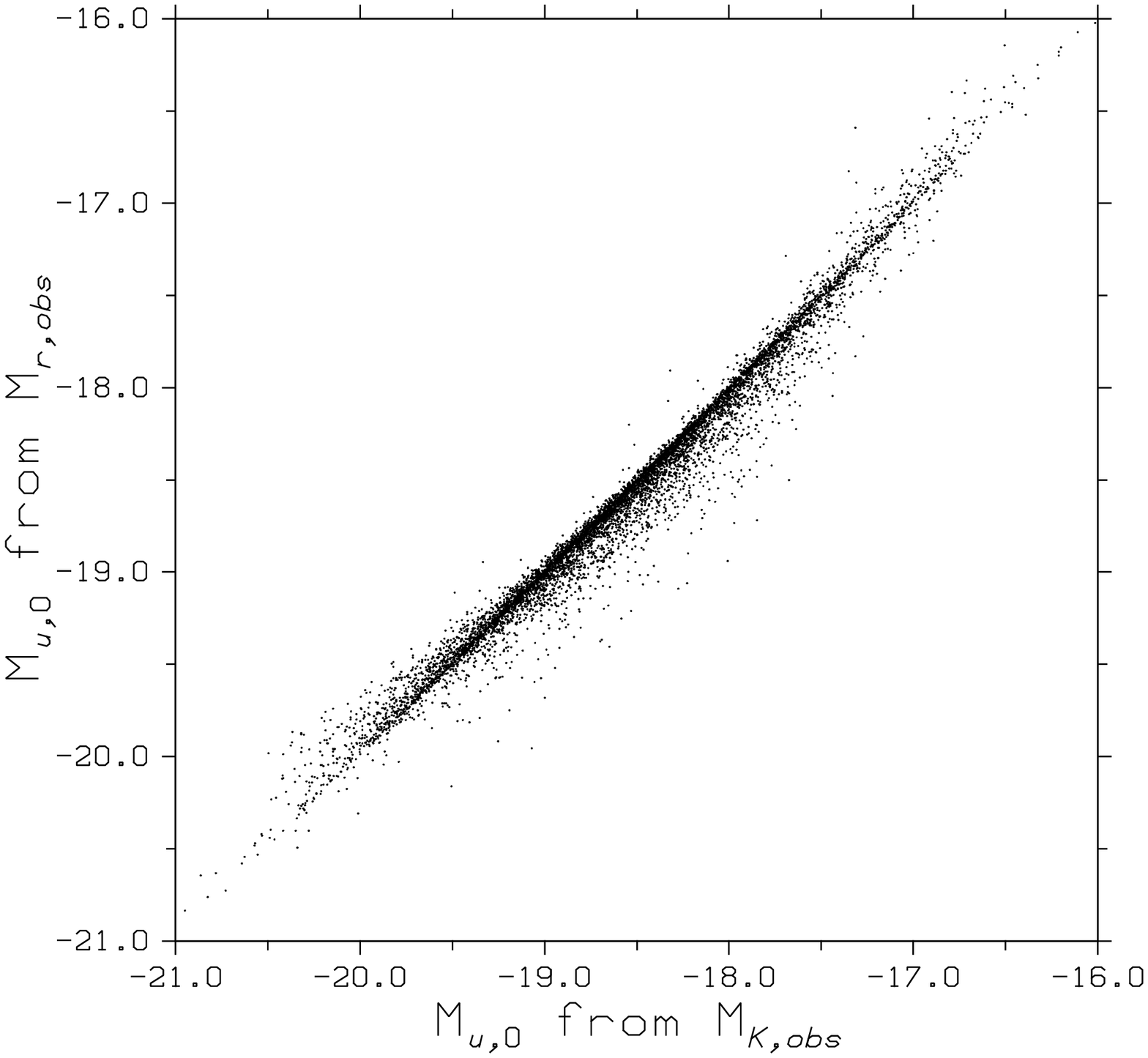}
\includegraphics[angle=-90,width=0.32\textwidth]{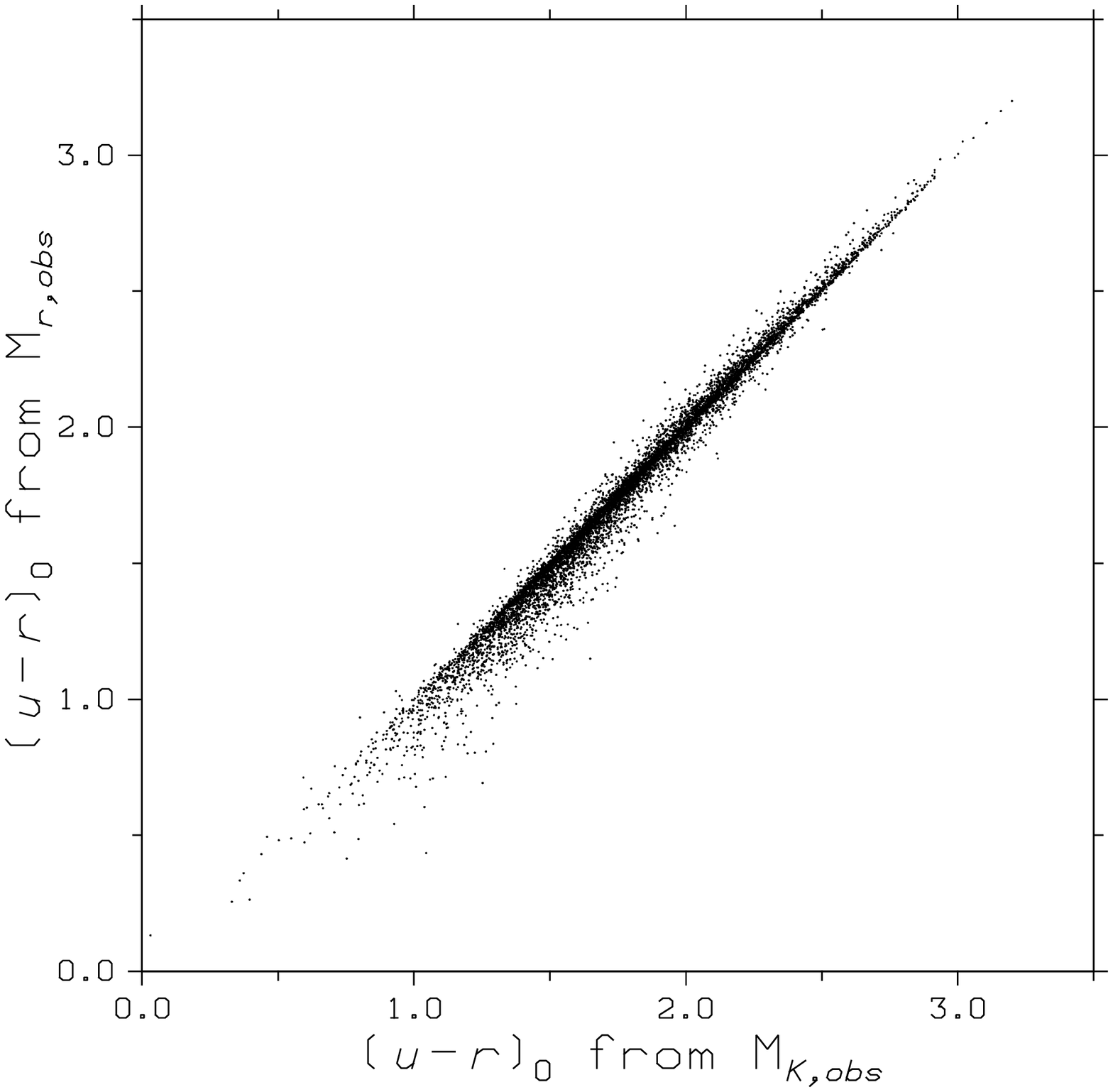}
\caption{ Comparisons between the inclination correction method using $\gamma(c,M_K)$ and
  $\gamma$'s in the shorter bands.
  The inclination correction using $\gamma(c,M_K)$ is
  straight forward.
  However, the use of $\gamma(c,M_{r,0})$ requires one more step
  (see Eq.~\ref{eq:RfromR}) to get $M_{r,0}$ first.
}
\label{fig_fromRorK}
\end{figure*}

\subsection{Obtaining $M_{r,0}$ from $M_{r,obs}$ and $c$}

We cannot observe the intrinsic $r$-band absolute magnitude $M_{r,0}$ directly.
But we can obtain $M_{r,0}$ by solving the following equation:
\begin{equation}
   M_{r,0}+ \gamma_{r-K}(c,M_{r,0}) \log_{10}(a/b) =M_{r,obs}, \label{eq:RR}
\end{equation}
where $M_{r,obs}$, $c$, and $a/b$ are observed quantities.
When we use a quadratic approximation as shown in Figure~\ref{fig_fitR},
then Eq.~\ref{eq:RR} becomes
\begin{equation}
   M_{r,0} + \Delta [(M_{r,0}+20.77)^2-1.10/0.223]=M_{r,obs},   
\end{equation}
where 
\begin{equation}
  \Delta \equiv 1.06\times0.223\times [1.35(c-2.48)^2-1.14]\log_{10}(a/b) \leq 0.
\end{equation}
Equivalently, we have
\begin{equation}
   (M_{r,0}+20.77) + \Delta [(M_{r,0}+20.77)^2-4.93]=(M_{r,obs}+20.77).
\end{equation}
The solution is
\begin{equation}
   M_{r,0}=-20.77+\frac{ -1 + \sqrt{ 1+4 \Delta (M_{r,obs}+20.77+4.93\Delta)} }
          { 2 \Delta },  \label{eq:RfromR}
\end{equation}
which gives the correct answer when $\Delta$ goes to zero.
This equation is valid for $[(M_{r,0}+20.77)^2-4.93]<0$ and $\Delta < 0$.
When $[(M_{r,0}+20.77)^2-4.93]\geq 0$ or $\Delta \geq 0$, we simply have 
$M_{r,0}=M_{r,obs}$.

Figure~\ref{fig_fromRorK} shows that extinction corrections using $M_{r}$ and $c$
give results quite close to those using $M_{K}$ and $c$.
In the left panel, we obtain the intrinsic r-band absolute magnitude using two methods.
For the x-axis, we make inclination corrections using
\begin{equation}
   \gamma_{r} \sim \gamma_{r-K} = \gamma_{r-K}(c,M_K),
\end{equation}
where $\gamma_{r-K}(c,M_K)$ is given in Table 1.
For y-axis, 
we make inclination corrections using Eq.~(\ref{eq:RfromR}).
The left panel shows that the two methods give a good agreement.
The middle and right panels are obtained similarly.
The $1\sigma$ error in $M_{r,0}$ derived from $M_{r,obs}$ and $c$ is given 
by the error in $\gamma_{r-K}$(c,$M_{r,0}$) times ${\rm log}(a/b)$ (see Eq. 14) 
or $0.208 {\rm log}(a/b)$ over the range from $M_{r,0}=-18.5$ to $-22.5$.

\subsection{Obtaining $M_{u,0}$ and $(u-r)_0$}

We can estimate $M_{u,0}$ from $M_{r,obs}$ and $c$.
First, we need to obtain $M_{u,0}$ as described in the previous subsection.
$M_{u,0}$ is 
\begin{equation}
 M_{u,0} = M_{u,obs}  - \gamma_{u-K}(c,M_{r,0}) \log_{10}(a/b),
\end{equation}
where $\gamma_{u-K}(c,M_{r,0})$ is given in Table 1.

We can also estimate $(u-r)_0$ from $a/b$ and $c$.
First, we need to obtain $M_{r,0}$.
Then, $(u-r)_0$ is 
\begin{equation}
 (u-r)_0 = (u-r)_{a/b}  - \gamma_{u-r}(c,M_{r,0}) \log_{10}(a/b),
\end{equation}
where $\gamma_{u-r}(c,M_{r,0})$ is given in Table 1.

\section{Dependence on \MakeLowercase{$u-r$} color}
In this paper, we have mainly considered dependence of $\gamma_{\lambda}$
on $K$ and $c$.
There may be more parameters that determine $\gamma_{\lambda}$.
In this section, we show that $u-r$ color can be a third independent parameter. 

\subsection{Slope vs. $(u-r)_0$}
We obtain the intrinsic color, $(u-r)_{0}$, as follows:
\begin{equation}
   (u-r)_0 =  (u-r)_{a/b}  - \gamma_{u-r}(c,M_{r,0}) \log_{10}(a/b),
\end{equation}
or,
\begin{equation}
   (u-r)_0 =  (u-r)_{a/b}  - \gamma_{u-r}(c,M_K) \log_{10}(a/b),
  \label{eq_ur0}
\end{equation}
where $\gamma_{u-r}(c,M_{r,0})$ and $\gamma_{u-r}(c,M_K)$ are given in Table 1.
In this section, we use Eq.~(\ref{eq_ur0}).

\begin{figure}[h!t]
\epsscale{1.0}
\plotone{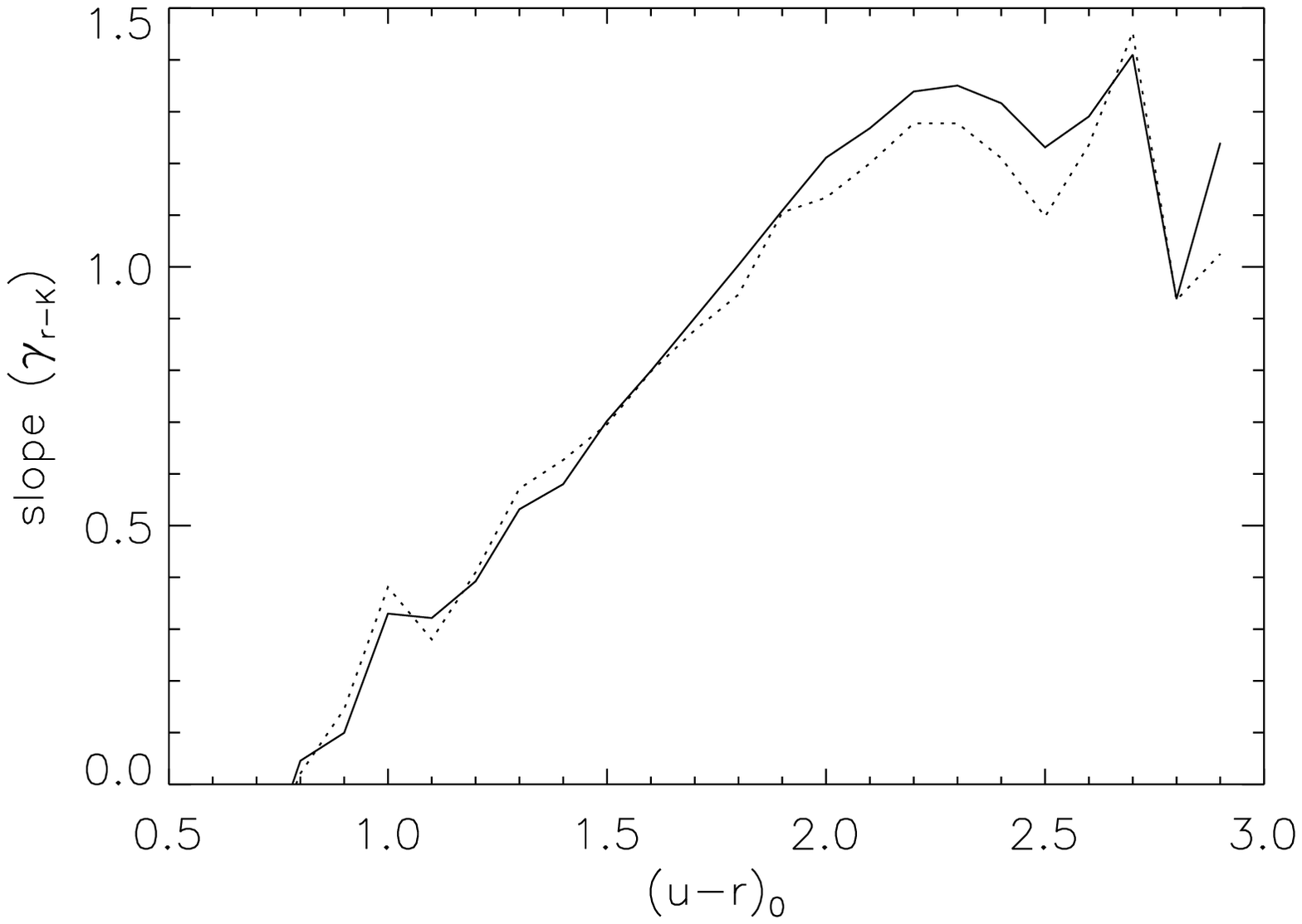}
\caption{ 
Dependence of $\gamma_{r-K}$ on $(u-r)_{0}$, where the subscript '$0$' denotes
  the inclination corrected value. 
  We use $\gamma_{u-r}=\gamma_{u-r}(c,M_K)$ (see Table 1) to derive
  the intrinsic color.
}
\label{fig_ur0}
\end{figure}

Figure~\ref{fig_ur0} shows that $\gamma_{r-K}$ strongly depends on $(u-r)_0$.
The Figure shows that the slope for $r-K_s$ is virtually zero when
$(u-r)_0$ is less than $\sim 1.0$.
The slope increases as  the color increases. 
It seems that the slope reaches a maximum at $(u-r)_0 \sim 2.5$.
The behavior of the slope is uncertain for $(u-r)_0 \gtrsim 2.5$.

\begin{figure}[h!t]
\plotone{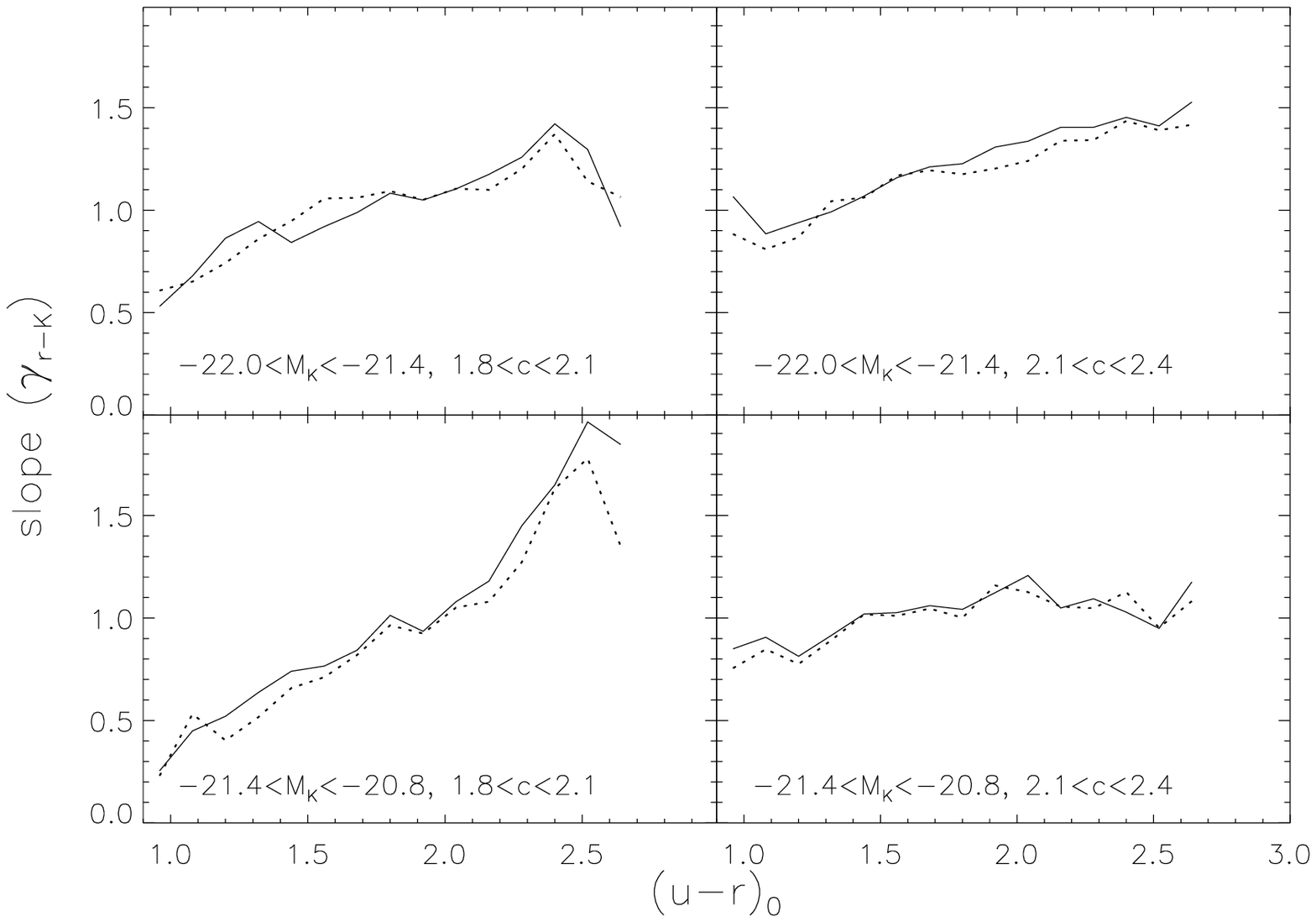}
\caption{
   $(u-r)_0$ as a third parameter.
   This figure shows that the slope depends not only on $c$ and luminosity
   but also the intrinsic $(u-r)$ color.
   Internal extinction is more significant when
   galaxies have larger intrinsic $(u-r)$ color.
}
\label{fig_3rd}
\end{figure}

\subsection{Dependence of $(u-r)_0$ on $c$ or \MakeUppercase{$M_K$}}
In this section, we investigate
the possibility of $(u-r)_0$ as a third parameter.
The third parameter should be independent of
the first 2 parameters (i.e. $c$ and $M_K$ in our case). 
However, Figure \ref{fig_2d_ur} shows that $(u-r)_0$, which 
corresponds to the y intercepts of the linear fits,
depends on the both $c$ and $M_K$.
Therefore, $(u-r)_0$ is not a completely independent parameter,
which means we cannot tell whether 
the change in slope observed in Figure~\ref{fig_ur0}
is entirely due to $(u-r)_0$ or just a different
realization of $c$ or $M_K$ dependency through the correlation between
them and $(u-r)_0$.

However, a careful look at Figure~\ref{fig_2d_ur} reveals that
$(u-r)_0$ varies between $\sim 1.5$ and $\sim 2.0$. 
One exception is the $(u-r)_0$ (i.e. the y intercept) in the
upper-right panel, which is as large as $\sim 2.4$.
{}From this observation, we can conclude that
the intrinsic scatter in $(u-r)_0$ itself is larger than
the systematic change in $(u-r)_0$ due to $c$ or $M_K$ change.
Therefore, if we properly limit the ranges of $c$ and $M_K$
and draw a plot similar to Figure~\ref{fig_ur0},
then we can tell whether or not $\gamma_{r-K}$ really depends on 
$(u-r)_0$.

For this purpose, we only consider galaxies in the following ranges:
$$
\begin{array}{l}
\mbox{group 1:  $-22.0 \leq M_K \leq -21.4$ and $1.8\leq c \leq 2.1$} \\
\mbox{group 2:  $-22.0 \leq M_K \leq -21.4$ and $2.1\leq c \leq 2.4$} \\
\mbox{group 3:  $-21.4 \leq M_K \leq -20.8$ and $1.8\leq c \leq 2.1$} \\
\mbox{group 4:  $-21.4 \leq M_K \leq -20.8$ and $2.1\leq c \leq 2.4$} \\
\end{array}
$$
Then, we study the dependence of $\gamma_{r-K}$ 
on $(u-r)_0$ for each group.

\subsection{$(u-r)_0$ color as a third parameter}
The discussion above indicates that the systematic change in $(u-r)_0$ in each group
due to changes in $c$ and $M_K$ should be marginal.
Figure~\ref{fig_3rd} shows the possibility that
$(u-r)_0$ can be a 3rd parameter:
For given $c$ and $M_K$, the slope $\gamma_{\lambda}$
shows dependence on $(u-r)_0$.
The general trend is that the slope is steeper, when
$(u-r)_0$ is larger. 
Note, however, that galaxies in group 4 do not show strong dependence on 
$(u-r)_0$, while those in group 3 show a very strong dependence.


\begin{figure*}
\includegraphics[angle=-90,width=0.90\textwidth]{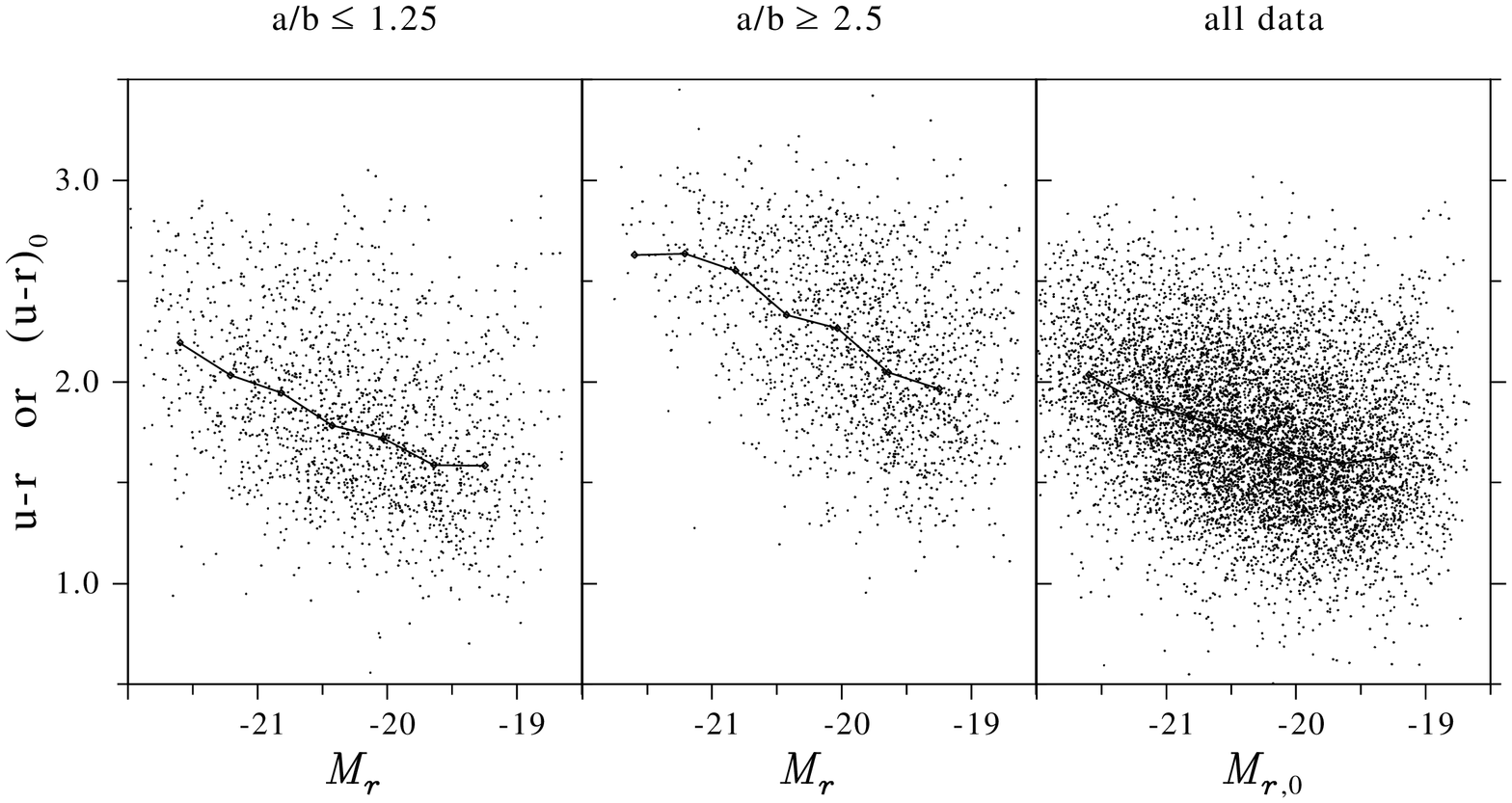}  
\caption{
Distributions of late-type galaxies in the color magnitude diagram. 
Galaxies with $a/b \leq$ 1.25 ($b/a \geq$ 0.8) are shown
in the left panel, and those with $a/b \geq$ 2.5 ($b/a \leq$ 0.4) are shown in the middle
panel. Distribution of all galaxies after inclination corrections are drawn
on the right panel.
The median color-magnitude relations are also drawn.
We find
that the group of highly inclined galaxies is significantly
shifted with respect to the almost face-on ones toward
fainter magnitudes and redder colors.
}
\label{fig_f12}
\end{figure*}

\section{Discussion}
Qualitatively speaking, our results are
consistent with earlier claims that the extinction
in spirals depends on luminosity (Giovanelli et al. 1995;
Tully et al. 1998).
Tully et al. (1998) derived an extinction law in $R$ band.
They found that galaxies with $M_{R}$ reaching 
$-21$ mag have $\gamma_R \sim 1.16$ and the amplitude of extinction
drops rapidly as luminosity decreases.
Similarly,
our Figure~\ref{fig_fitR} shows that 
$\gamma_{r-K} \sim 1.1$ for $M_r\sim -21$ and it declines
rapidly with decreasing luminosity.
However, there are also differences. For example,
$\gamma_R$ in Tully et al.~(1998) reaches as large as $\sim 1.4$
for galaxies brighter than -21 mag in the $R$ band.
We do not observe such a rise of $\gamma_r$ in our results.
However, since the definition of $r$-magnitude is different and 
$\gamma_r$ depends on other parameters, such as $c$ and $u-r$,
it is very difficult to understand the origin of this discrepancy.

    Shao et al.~(2007) found $\gamma_r \sim 1.37$ when they used
    the $r$-band axis ratio, which is substantially larger than
    our values. The difference may stem from the
    difference in fitting methods. We studied how $r-K_s$ color
    changes as a/b changes. Therefore the value of $\gamma$ in our
    study is actually $\gamma_{r-K}$, not $\gamma_r$. 
    The value of $\gamma_r$ will be given by
    $\gamma_r \sim \gamma_{r-K} + \gamma_K \sim 1.15 + 0.26 \sim 1.41$,
    where $1.15$ is the average value of $\gamma_{r-K}$
    near the maximum (see Figure 5) and $0.26$ is the estimated
    $\gamma_{k}$ (Masters et al.~2003). The definition of
    $a/b$ is also different. Our $a/b$ is the $i$-band isophotal
    axis ratio, while that of Shao et al.~(2007) is the $r$-band axis ratio
    obtained from the best fit of the images of galaxies with an 
    exponential profile convolved with the point spread function.

On the other hand, there are several observations in $I$ band.
Giovanelli et al. (1994) found the extinction
in $I$ band of
$\gamma_I\sim 1.05 \pm 0.08$. Masters et al. (2003) found that
$\gamma_I\sim 0.94$.
Therefore, our result of $\gamma_r \sim 1.1$ agrees with the
common wisdom: there is more extinction in shorter passband.

The concentration index, $c$, is related to morphology of late-type galaxies.
Earlier studies show that, on average, 
early late-type galaxies have higher concentration
(see, for example, Shimasaku et al. 2001; Strateva et al. 2001; 
Goto et al. 2003; Yamauchi et al. 2005). 
Therefore dependence of inclination effects on the concentration index implies
that galaxy morphology is an important factor, which is consistent
with earlier findings (see, for example, de Vaucouleurs et al. 1991; Han 1992).
It is interesting that internal extinction is maximum when $c\sim 2.5$
which corresponds to Sb galaxies (see Shimasaku et al. 2001).

Choi et al. (2007) have shown that the sequence of late-type galaxies
in the color-magnitude diagram can be significantly affected by
the internal extinction. When the sequences are compared between
late types with $b/a>0.8$ (nearly face-on) and $b/a<0.4$ (nearly edge-on),
a very large departure is observed for the sequence of the inclined 
late-type galaxies relative to that of the face-on late types as the 
inclined galaxies appear fainter and redder. The degree of departure
was found to be maximum for intermediate luminosity late types with
$M_r \approx -20.5+5{\rm log}h = -21.1$. No such extinction effect was
found for early-type galaxies. It will be very interesting to see
if our extinction-correction prescription restores the color-magnitude
diagram of inclined late-type galaxies.

In Figure~\ref{fig_f12}, we plot our late-type galaxies 
in the color magnitude diagram. 
The left panel of Figure~\ref{fig_f12} shows the galaxies with
$a/b \leq$ 1.25 ($b/a \geq$ 0.8) and the middle shows those
with $a/b \geq$ 2.5 ($b/a \leq$ 0.4).
We find that the group of highly inclined galaxies is significantly
shifted with respect to the almost face-on ones toward fainter
magnitude and redder colors. 
The result in Figure~\ref{fig_ur0} indicates that
the color of very blue galaxies is less affected by 
the internal extinction.
However, Figure~\ref{fig_ur0} does not tell much about color change
of very red galaxies (i.e. redder than $u-r \sim$ 2.5) due to
lack of data.
Note that Choi et al. (2007) observed that both very blue and very red galaxies
are less effected by the internal extinction.
The right panel of Figure~\ref{fig_f12} is the inclination-corrected  
color magnitude diagram. 
The inclination correction is done based on $M_K$- and $c$-dependence 
of $M_r$ and $u-r$. 
To be more specific, we use Eqs. (13) and (21) to obtain $M_{r,0}$ and $(u-r)_0$. 
We do not draw the reddening vectors (caused by internal extinction)
on the color-magnitude diagram becase, due to $c$-dependency,
infinite number of reddening vectors are possible for a given value of ($M_r$, $u-r$).
If one wish to draw `average' reddening vectors for a fixed value of $\log(a/b)$, one can do that
by using $\gamma_{u-r}(M_{r,0})$ and $\gamma_{r}(M_{r,0})$ in Table 1 (or 2).
Of course, to get the $\gamma$ values, we first need to obtain $M_{r,0}$
from Eq.~(\ref{eq:RfromR}). 

\section{Conclusion}
Our major conclusions are as follows.
\begin{enumerate}
\item{} We have shown that the slope $\gamma_{\lambda}$ 
depends on both $K_s$-band absolute magnitude $M_K$ and the concentration index
$c$. The fitting functions for the relations are given in Table 1 and 2.
\item{} We have also shown that the slope $\gamma_{\lambda}$ 
depends on both the concentration index $c$ and $r$-band absolute
magnitude $M_{r,0}$, where
the subscript `$0$' denotes the value after inclination correction.
The relations are also given in Table 1 and 2.
\item{} We have derived analytic formulae giving the extinction-corrected 
      magnitudes from the observed optical parameters (see, for example,
      Equation (\ref{eq:RfromR})).
\item{} We have shown that $(u-r)_0$ can be a third parameter
that determines the slope $\gamma_{\lambda}$.
\end{enumerate}

\acknowledgments
The authors thank Dr. Yun-Young Choi for her collaborative work 
on the volume-limited galaxy sample that this work used.
J.Y.C.'s work was supported by the Korea Research Foundation grant
funded by the Korean Government (KRF-2006-331-C00136). 
C.B.P. acknowledges the support of the Korea Science and Engineering
Foundation (KOSEF) through the Astrophysical Research Center for the
Structure and Evolution of the Cosmos (ARCSEC).

Funding for the SDSS and SDSS-II has been provided by the Alfred P. Sloan
Foundation, the Participating Institutions, the National Science 
Foundation, the U.S. Department of Energy, the National Aeronautics and
Space Administration, the Japanese Monbukagakusho, the Max Planck
Society, and the Higher Education Funding Council for England.
The SDSS Web Site is http://www.sdss.org/.

The SDSS is managed by the Astrophysical Research Consortium for the
Participating Institutions. The Participating Institutions are the
American Museum of Natural History, Astrophysical Institute Potsdam,
University of Basel, Cambridge University, Case Western Reserve University,
University of Chicago, Drexel University, Fermilab, the Institute for
Advanced Study, the Japan Participation Group, Johns Hopkins University,
the Joint Institute for Nuclear Astrophysics, the Kavli Institute for
Particle Astrophysics and Cosmology, the Korean Scientist Group, the 
Chinese Academy of Sciences (LAMOST), Los Alamos National Laboratory, 
the Max-Planck-Institute for Astronomy (MPIA), the Max-Planck-Institute 
for Astrophysics (MPA), New Mexico State University, Ohio State University,
University of Pittsburgh, University of Portsmouth, Princeton University,
the United States Naval Observatory, and the University of Washington.

\clearpage
\begin{deluxetable}{llllll}  
\tabletypesize{\footnotesize}
\tablecaption{Fitting Functions (for most probable values)}
\tablewidth{0pt}
\tablehead{
   \colhead{} & 
   \colhead{Dependence on c} & 
   \colhead{Dependence on $M_K$} & 
   \colhead{Dependence on $M_{r,0}$} & 
   \colhead{$M_K$ \& c} & 
   \colhead{$M_{r,0}$ \& c} \\
   \colhead{} &
   \colhead{$\gamma(c)$\tablenotemark{a}} &
   \colhead{$\gamma$($M_K$)\tablenotemark{b}} &
   \colhead{$\gamma$($M_{r,0}$)\tablenotemark{c}} &
   \colhead{$\gamma$(c,$M_K$)} &
   \colhead{$\gamma$(c,$M_{r,0}$)}
}
\startdata 
$r$-$K_s$  &
-1.35(c-2.48)$^2$+1.14 &
-0.089($M_K$+22.9)$^2$+1.16  &
-0.223($M_{r,0}$+20.8)$^2$+1.10  &
1.02$\gamma(M_K) \gamma(c)$ &
1.06$\gamma(M_{r,0}) \gamma(c)$  \\
$u$-$K_s$  &
-3.37(c-2.57)$^2$+2.61 &
-0.187($M_K$+22.9)$^2$+2.24 &
-0.318($M_{r,0}$+20.8)$^2$+2.04  &
0.48$\gamma(M_K) \gamma(c)$   &
0.52$\gamma(M_{r,0}) \gamma(c)$ \\
$u-r$  &
-2.02(c-2.63)$^2$+1.49 &
-0.098($M_K$+22.9)$^2$+1.09  &
-0.095(M$_{r,0}$+20.9)$^2$+0.94  &
0.94$\gamma(M_K) \gamma(c)$   &
1.03$\gamma(M_{r,0}) \gamma(c)$  \\
             \hline
\enddata
\tablenotetext{a}{Range of validity: $1.74 \leq c \leq 3.06$}
\tablenotetext{b}{Range of validity: $-23.25 \leq M_K \leq -19.95$}
\tablenotetext{c}{Range of validity: $-21.95 \leq M_{r,0} \leq -18.65$}
\tablenotetext{*}{Note: Expressions for $\gamma_u$ can be obtained by
   $\gamma_u=\gamma_{u-K}-\gamma_K \sim \gamma_{u-K}$, where we assume 
   $\gamma_K$ is small. Earlier studies (e.g.~Tully  et al.~1998;
   Masters et al.~2003) showed $\gamma_K \lesssim 0.2$.
   Similarly, we have $\gamma_r=\gamma_{r-K}-\gamma_K \sim \gamma_{r-K}$.}

\label{table_max}
\end{deluxetable}
\begin{deluxetable}{llllll}  
\tabletypesize{\footnotesize}
\tablecaption{Fitting Functions (for the median)}
\tablewidth{0pt}
\tablehead{
   \colhead{} & 
   \colhead{Dependence on c} & 
   \colhead{Dependence on $M_K$} & 
   \colhead{Dependence on $M_{r,0}$} & 
   \colhead{$M_K$ \& c} & 
   \colhead{$M_{r,0}$ \& c} \\
   \colhead{} &
   \colhead{$\gamma(c)$\tablenotemark{a}} &
   \colhead{$\gamma$($M_K$)\tablenotemark{b}} &
   \colhead{$\gamma$($M_{r,0}$)\tablenotemark{c}} &
   \colhead{$\gamma$(c,$M_K$)} &
   \colhead{$\gamma$(c,$M_{r,0}$)}
%
}
\startdata 
$r$-$K_s$  &
-1.13(c-2.49)$^2$+1.08 &
-0.072($M_K$+23.1)$^2$+1.13  &
-0.215($M_{r,0}$+20.8)$^2$+1.08  &
1.07$\gamma(M_K) \gamma(c)$   &
1.09$\gamma(M_{r,0}) \gamma(c)$  \\
$u$-$K_s$  &
-2.84(c-2.61)$^2$+2.49 &
-0.154($M_K$+23.1)$^2$+2.21  &
-0.296($M_{r,0}$+20.9)$^2$+1.99  &
0.52$\gamma(M_K) \gamma(c)$  &
0.55$\gamma(M_{r,0}) \gamma(c)$  \\
$u-r$  &
-1.70(c-2.69)$^2$+1.43 &
-0.082($M_K$+23.1)$^2$+1.08  &
-0.081($M_{r,0}$+21.2)$^2$+0.92  &
1.02$\gamma(M_K) \gamma(c)$   &
1.14$\gamma(M_{r,0}) \gamma(c)$ \\  
             \hline
\enddata
\tablenotetext{a}{Range of validity: $1.74 \leq c \leq 3.06$}
\tablenotetext{b}{Range of validity: $-23.25 \leq M_K \leq -19.95$}
\tablenotetext{c}{Range of validity: $-21.95 \leq M_{r,0} \leq -18.65$}
\tablenotetext{*}{Note: Expressions for $\gamma_u$ can be obtained by
   $\gamma_u=\gamma_{u-K}-\gamma_K \sim \gamma_{u-K}$, where we assume 
   $\gamma_K$ is small. Earlier studies (e.g.~Tully  et al.~1998;
   Masters et al.~2003) showed $\gamma_K \lesssim 0.2$.
   Similarly, we have $\gamma_r=\gamma_{r-K}-\gamma_K \sim \gamma_{r-K}$.}
             
\label{table_max}
\end{deluxetable}
\end{document}